\documentstyle[12pt,aas2pp4]{article}
\input{epsf.sty}
\lefthead{Henriksen and Cousineau}
\righthead{An X-ray Survey of Paired Galaxies}

\begin{document}

\title{An X-ray Survey of Galaxies in Pairs}

\author{Mark Henriksen and Sarah Cousineau}
\affil{Physics Department, University of North Dakota,
    Grand Forks, ND 58202-7129}

\begin{abstract}

Results are reported from the first survey of 
X-ray emission from galaxies in pairs.
The sample consists of fifty-two pairs of galaxies from the 
Catalog of Paired Galaxies (CPG) 
Karachentsev (1972) whose coordinates overlap 
{\it ROSAT} Position Sensitive Proportional Counter (PSPC) 
pointed observations. 
The mean observed log l$_{x}$ for early-type pairs is 41.35$\pm$0.21
while the mean log l$_{x}$ predicted using the l$_{x}$-l$_{b}$
relationship for isolated early-type galaxies 
is 42.10$\pm$0.19. With 95\% confidence, the galaxies
in pairs are underluminous in the X-ray, compared to isolated galaxies,
for the same l$_{b}$.
A significant fraction of the mixed pair sample also appear similarly
underluminous.
A spatial analysis shows that the X-ray emission from
pairs of both types typically has an extent of $\sim$10 - 50 kpc,
much smaller than group intergalactic medium and thus likely
originates from the galaxies. 
CPG 564, the most X-ray luminous early-type pair, 
4.7$\times$10$^{42}$ ergs
sec$^{-1}$, is an exception. 
The extent of it's X-ray emission, $>$169 kpc, and
HWHM, $\sim$80 kpc, is comparable
to that expected from an intergalactic medium. 
The sample shows only a weak correlation, $\sim$ 81\% 
confidence, between  l$_{x}$ and l$_{b}$, presumably due
to variations in gas content within the galaxies.
No correlation between
l$_{x}$ and the pair velocity difference ($\Delta$v), separation $\Delta$r,
or far-infrared luminosity (l$_{fir}$) is found though the
detection rate is low, 22\%.

\end{abstract}


\section{Introduction}

Studies of paired galaxies in the radio, infrared, and optical
provide important details regarding the
role of interactions and mergers 
in the evolution of galaxies. Evidence of interaction is inferred from
excess emission in the far-infrared (FIR) and optical relative to isolated galaxies which
indicates an increased level of star formation from the interaction (Bushouse 1987).
While these data provide a description of
the star formation history of the galaxies, X-ray
studies address the fate of gas heated
by the large energy input from stellar evolution, such
as the formation of a superwind (Heckman, Armus, and Miley 1990). Comparisons
of the X-ray and optical luminosity of paired with 
isolated galaxies provides important
details of the connection between star formation
and hot gas. Studies of the hot, X-ray emitting
gas content of interacting galaxies also has
important implications for the enrichment of the intracluster
and intergalactic medium.
Recent studies of cluster abundances indicate that the enrichment of
the ICM has changed little since z$\sim$0.3 suggest that enrichment
must have occurred in protogalaxies rather than during cluster evolution
(Mushotzky and Loewenstein 1997). Though local systems are studied here, the
results pertaining to the fate of hot gas in paired, interacting galaxies
could be applied equally well to protogalactic pairs within clusters.

The X-ray observations are also in a unique position to 
address the  connection between pairs and groups of
galaxies. Groups typically
have a hot intergalactic medium. If
mixed- and early-type pair morphologies were merger
remnants of groups, then they would retain 
the hot intergalactic medium of the group, being
characterized by an extended X-ray component of $>$200
kpc, as is typical of groups (Mulchaey et al. 1996). In addition, the 
total X-ray luminosity of compact groups 
is higher than that expected from the
galaxies themselves (Ponman and Bourner 1997). Detection
of the fossil intergalactic medium would make
a strong argument that mergers take place within groups.

In this paper, we present the characteristics of the sample and details
on the data analysis in section 2. In section 3 we present the results of
correlation tests on the entire sample (3.1) 
and comparisons of the X-ray emission of paired versus
isolated galaxies for spiral pairs
(3.2), early-type pairs (3.3), and mixed pairs (3.4). 
In section 4, results of the
spatial analysis are presented. In section 5,
the results are summarized. A value of
50 km sec$^{-1}$ Mpc$^{-1}$ is used for all calculated quantities.

\section{Data Analysis}

We have checked the {\it ROSAT} PSPC 
pointed data archives for serendipitous occurrences
of over six hundred pairs of galaxies from the optically 
selected Catalog of Paired Galaxies (CPG) by
Karachentsev (1972). Fifty-eight
pair positions overlap PSPC pointings. 
The PSPC is the
best instrument
to use for this survey because of its larger field of view and greater
sensitivity when compared to other available instruments such
as the
{\it ROSAT} high resolution imager or the {\it EINSTEIN} imaging 
proportional counter (IPC). The 
pointed observations, while
not providing all-sky coverage, have the advantage of much longer
exposure times than those of the
{\it ROSAT} all-sky survey (RASS), $\sim$550 seconds, which is necessary 
since most of these sources are below the flux limit of the
RASS,$\sim$4$\times$10$^{-13}$ ergs cm$^{-2}$ sec$^{-1}$.
Using {\it ROSAT} for
a survey of optically selected groups, Mahdavi et al. (1997) 
found that only seven out of thirty-two objects are detected on RASS fields 
of which three are intrinsically much brighter
Abell clusters. We detect 11 out of 49 pairs, excluding
AGN, at
$>$ 3$\sigma$. The range in X-ray luminosity  of
groups in their survey exceeds more than 50\% of the
pairs so that our higher detection rate
must be attributed to the longer pointed observations. 
For a velocity of 6000 km sec$^{-1}$, typical of the paired galaxies,
the minimum measurable luminosity with RASS would be
$\sim$3$\times$10$^{41}$ ergs
sec$^{-1}$. Thus, using the correlations found in this
paper, we could expect to detect only the X-ray brightest 
pairs, typically those containing Seyfert galaxies,
which are not suitable for study of the hot gas content.

We excluded pairs with have high velocity
difference and are therefore chance alignments:
CPG 391 ($>$4000 km s$^{-1}$), CPG 321 ($>$500 km s$^{-1}$), and CPG 166 
($>$2000 km s$^{-1}$)). Pairs 
in galaxy clusters were also eliminated since they
are contaminated by the intracluster medium: 
CPG 588 (in A2634), CPG 464
(in A2063), CPG 361 (in Coma). CPG 353 is in the outskirts of the Virgo cluster
where the contribution from the intracluster medium is
negligible and it is included in the
sample. The best fit temperature for CPG 353 is 0.82 keV, more 
typical of emission associated with galaxy
than cluster. 
Since we are addressing the nature of the hot gas content in
galaxy pairs, those with AGN were eliminated:
CPG 510, CPG 419, CPG 238. CPG 234 also 
has an extended component (2.3' in extent or 40 kpc) as well as
an AGN (section 4)
and is included in the analysis. The three pairs with AGN are included
in the data Tables and shown in the figures but are not included in
any analysis.

Four of the detected
pairs were {\it ROSAT} targets: 125, 218, 278, and 353. 
Two are spiral pairs and two are mixed-type pairs (CPG 278 and CPG
353).
Therefore, the collection of 8 early-type pairs contains only serendipitous
observations and represents the unbiased X-ray characteristics
of the sample.

Each galaxy pair was first located on the Digitized
Sky Survey (DSS) to check the accuracy of the CPG positions. 
The right ascension
and declination given in Tables 1-3 have been modified for some pairs
to reflect more accurate coordinates obtained from the DSS
than in the CPG. The region size 
of the image used to measure the X-ray emission 
was chosen with consideration of the nominal extent of possible
extended X-ray emission
associated with the galaxies themselves including 
plumes (Dahlem et al. 1996; Fabbiano 1988), halos (Trinchieri and
Fabbiano 1985; Forman, Jones, and Tucker 1985) or the nominal
core radius of intergalactic emission
from a fossil group using NGC 2300 as a prototype (Davis et al. 1996). It is important to use a fixed, distance-independent
sized region for all
pairs. This region is a circle with radius equal
to the pair separation (in kpc) plus twice the nominal galaxy radius, $\sim$50 kpc. 
In most cases, the
emission from either galaxy and from an intergalactic medium
would be included within this region. Contaminating point sources
were removed when necessary. For wider pair separations or especially
difficult locations on the image, 
a circular region of radius 50 kpc 
centered on the coordinates of the
optically brightest galaxy was used. The size and location of
each source region is given in Table 4. Standard
PSPC analysis procedures for extended sources, 
including
background subtraction, vignetting and exposure map
 corrections (Table 4), and charged particle elimination, were
 followed as outlined in
Snowden et al. (1994) to
analyze the source regions using routines which are part
of the imaging and spectral analysis package
in PROS version 2.4 (Deponte, J. et al. 1995).
 
Signal-to-noise ratios are given in Table 4.
The detection threshold is 3$\sigma$
or higher significance. For those source regions with count rates
below the detection threshold, an upperlimit is
calculated equal to 3$\sigma$. 
Spectra for all detected objects were modeled
using XSPEC version 9.0 (Arnaud 1996) to obtain
the flux. The model consists of a Raymond
and Smith plasma code (1977) with
absorption cross sections (Morrison and McCammon 1983) using
galactic column densities at the source positions (Stark 1992) (values
given in Table 4), and
solar abundance ratios (Anders and Grevesse 1989).
The data was grouped to contain at least twenty counts in each bin to
increase the reliability of the results obtained from a chi-square fit
(Nousek and Shue 1989). The best-fit plasma temperatures
and 90\% confidence range  from the spectral analysis are given
in Table 5.

Upperlimts are converted to fluxes
using a Raymond and Smith model convolved with the PSPC
response matrix
assuming solar abundance and a temperature of 1 keV for
early-type and mixed pairs (Kim, Fabbiano, and
Trinchieri 1992), for which the X-ray emission 
is generally dominated
by the elliptical galaxy,
and 5 keV for spiral pairs (Kim, Fabbiano, and
Trinchieri 1992).
{\it ROSAT} and Advanced Satellite for Astrophysics and
Cosmology ({\it ASCA}) spectra of spiral
galaxies show a soft component in addition to the hard
component which dominates the {\it EINSTEIN} spectra (see references
in section 3) in general agreement with the best-fit values
reported in Table 5 for spiral pairs. 
In the most extreme case, using a temperature of $\sim$0.4 keV from 
{\it ROSAT} spectra of starburst 
galaxies rather
than the nominal temperature would increase the luminosities in Tables 1-3 by a factor
of only $\sim$1.3. Since the upperlimits for the spiral pairs are
not very constraining in the correlations,
the exact temperature used to convert the upperlimits is not critical.
The distance is derived
from the average of the galaxy pair velocities and
used in the luminosity calculation.

X-ray, optical, and FIR luminosities 
and basic data are shown in Table 1-3 for
spiral, early-type, and mixed pairs, respectively.
The content of the tables are the following:
column (1) is the pair number in the CPG,
column (2) is the galaxy classification from NASA/IPAC Extragalactic
Database (NED), column (3) and (4) are the RA and DEC
in J2000 coordinates, column (5) is
the separation of galaxies in minutes of arc,
column (6) is the galaxy recessional velocity in km sec$^{-1}$, 
column (7) is the blue luminosity in units of 10$^{43}$ ergs sec$^{-1}$,
column (8) is the background subtracted X-ray count rate 
in counts sec$^{-1}$, column (9) is the 1$\sigma$
error in counts sec$^{-1}$,
column (10) is the X-ray luminosity in the 0.25-2 keV band in units
of 10$^{41}$ ergs sec$^{-1}$, column (11) is l$_{FIR}$. Data in columns (1,3-7) are based
on information from Karachentsev (1972). 
A Hubble constant of 50 km sec$^{-1}$ Mpc$^{-1}$ is used to
calculate the luminosities.
Column 11 is calculated from fluxes obtained 
from the Catalogue
of Galaxies and Quasars Observed in the IRAS Survey, Version 2.1
as well as the Point Source Catalog and the Faint Source Catalog. 
The FIR sources occur within a 25 kpc search
radius centered on each of the galaxy positions. 

Table 4 contains supplemental information to that in Tables 1-3. 
The content of the tables are the following:
column (1) is the pair number in the CPG,
column (2) is the PSPC sequence number, 
column (3) is the size of the source region in arc min,
column (4) is the
distance of the analyzed region from the center of the image in arcmin, 
column (5) is
the exposure time, column (6) is the signal-to-noise/3, column (7)
is the exposure map correction which includes vignetting
and correction for the instrument supports, column (8) is
the hydrogen column density,
column (9) are galaxy UGC, NGC, IC, and CGCG designations
of the galaxies in Tables 1-3.

\section{Results}

\subsection{Correlations}

Normal galaxies all show a correlation
between the optical and X-ray luminosity.
These correlations allow linking the stellar sources
to the amount of hot gas, both stellar and diffuse, 
thus providing
clues to the origin of the X-ray emission. 
Correlation tests
were performed on the subsample of pairs, which included 
all galaxy types,
using l$_{x}$ as the dependent variable 
and the physical quantities: 
l$_{b}$, l$_{FIR}$, $\Delta$r, and $\Delta$v as independent variables. 

The l$_{x}$-l$_{b}$ correlation and other correlation tests were performed using
ASURV Rev. 1.2 (LaValley, Isobe, and Feigelson 1992) which
implements the methods described in Isobe, Feigelson, and Nelson (1986).
Table 6 summarizes the results and has the following column headings
and uses the abbreviations which are given in parentheses.
Column (1) (labeled DV) is 
the dependent variable in the bivariate correlation
tests. Column (2) (labeled IV) is the independent
variable. Column (3) (labeled DET/TOT) is the number of detections and total number of observations used in the 
correlation test, Column (4) (labeled GKT) is the probability
of a correlation using
the Generalized Kendall's Tau. Columns (5-8) are the slope and intercept,
with associated errors, from a linear regression analysis using the EM algorithm.

The mean values and variances given in Table 7 are obtained using the non-parametric
Kaplan-Meier estimator 
	(Babu and Feigelson 1996). 
Upperlimits are included in both the paired
	and isolated galaxy samples.
This approach is valid as long as censoring 
	in the sample is random. 
	Visual
	inspection of the various samples shows the condition of random
	censoring is met. 

The first variables tested were l$_{x}$ and l$_{b}$; the data are shown in
Figure 1. 
The range in l$_{b}$
is narrow compared to normal galaxies, a factor of less than 100, and very high, with a mean value
$\sim$ 10$^{44}$ ergs sec$^{-1}$. The range in log (l$_{x}$) is much larger, a factor of
$\sim$10$^{4}$, and the mean is lower, 40.86 (see table
7). 
There
is a general linear trend between log l$_{x}$ and log l$_{b}$. However, 
the scatter is significant at all luminosities and increases
at higher l$_{b}$. The figure also shows that the l$_{x}$-l$_{b}$
locus is very different for each pair type.  
Even though there is a weak correlation present, 81\%, it is not very tight as
is apparent in the large error in the slope.

Strong FIR emission is an indicator of active star 
formation since interacting galaxies show a high l$_{FIR}$/l$_{b}$.
Close pairs of galaxies with spirals are generally stronger FIR emitters 
(Xu and Sulentic 1991) implying that they are interacting and 
should be emitting
X-rays associated with massive stars, X-ray binaries, 
and interstellar gas heated by
Type II supernovae. Thirty-six 
of the spiral pairs and mixed pairs 
were detected
in the FIR but 
none of the early-type
pairs were detected. Figure 2 shows l$_{FIR}$ and l$_{x}$.
The quantity l$_{FIR}$, is calculated from the cataloged
FIR flux:

Log(F$_{FIR}$ = Log(1.26$\times$(F60 + F100)), where F60 = 2.58$\times$10$^{-14}$$\times$
F$_{\nu}$(60$\mu$) and F100 = 1.00$\times$10$^{-14}$$\times$
F$_{\nu}$(100$\mu$). The fluxes at 60$\mu$ and 100$\mu$ are in Janskys.

We ran correlation tests
between l$_{x}$-l$_{FIR}$ and l$_{x}$-${\Delta r}$ for the pair samples
which have 36 and 49 pairs, respectively.
The correlation test results are shown in Table 6; 
no correlation is present in either instance.

Figure 3 shows 
X-ray luminosity versus galaxy separation. Though the
eye sees a trend of decreasing X-ray luminosity
with decreasing separation, they are only weakly correlated (69\%).

\subsection{Spiral Galaxies}

Twenty-five pairs of spiral galaxies are
analyzed. The X-ray luminosities
range from $<$2.6 x 10$^{40}$ to 10$^{42}$ ergs sec$^{-1}$.
The l$_x$,l$_b$ for pairs is shown in Figure 4 along with
l$_x$,l$_b$ for a sample of normal 
spiral galaxies taken from the sample of galaxies observed with {\it EINSTEIN} 
(Fabbiano, Gioia, and Trinchieri 1988; Trinchieri and Fabbiano 1985).
The data are shown in the 0.5 - 3 keV band. 
Galaxies classified as T=0,10 (very early type or irregular) 
were removed to get a representative sample of normal spirals.
The pairs
are converted from the {\it ROSAT} band using the canonical
spectrum discussed in section 2 to derive luminosities. A value
of H$_{0}$ = 50 km sec$^{-1}$ Mpc$^{-1}$is used in both samples.
L$_{x}$ is linearly correlated with l$_b$ on a log-log scale for
the normal galaxy sample,
$>$ 99.9\% confidence. We found 
the mean log (l$_{x}$) of the paired spirals
to be 40.82 $\pm$0.11 ergs sec$^{-1}$, 
higher than the mean value, 40.06 $\pm$0.06 ergs sec$^{-1}$,
predicted using the l$_{x}$-l$_{b}$
relationship for isolated
spirals observed with {\it EINSTEIN}.

Peace and Sansom (1996) analyzed  
approximately 20 normal spiral galaxies observed with
the PSPC. The galaxies in their sample have m$_{B}$ $<$ 13
and PSPC exposures $>$ 13,000 seconds. 
They found a mean luminosity of $\sim$10$^{40}$. This
mean is higher than that found for the entire sample
observed with {\it EINSTEIN} but similar to the value reported
above for the subsample of optically bright galaxies. Using
hardness ratios, these authors found that the PSPC spectra are
consistent with a temperature of 0.85 keV and a range
of 0.17 - 3.72 keV, suggesting a softer component may
also be present in addition to the $>$3 KeV hard component
(Kim, Fabbiano, and Trinchieri 1992). {\it ASCA} observations
of the spiral galaxy, NGC 2903 (Mizuno et al. 1996),
also show a hard, $\sim$ 5 keV component and a soft $\sim$ 0.4 keV component
similar to the spectra they find for the starburst galaxies
M82 and NGC 253. We performed a spectral analysis 
of all of the detected pairs.
For most of the spiral galaxies, the data only provide lower limits
to the temperature when fitting a Raymond and
Smith model (see Table 5). 
Only for the spiral pair CPG 218, consisting of M81 and M82, is the
90\% temperature range narrow, 0.7 - 1.1 keV.
The best-fit temperatures
are consistent with a softer
X-ray component dominating the {\it ROSAT} emission
as compared with {\it EINSTEIN}. Thus, comparing to the
sample in Peace and Sansom, the mean for spiral pairs,
log l$_{x}$ = 40.82 is significantly higher.

Similar conclusions resulted from a comparison of
{\it EINSTEIN} observations of a subsample of normal
galaxies with a sample of peculiar, blue galaxies.  The sample of irregular
galaxies had several galaxies
with much higher l$_{x}$/l$_{b}$ than the normal galaxies (Fabbiano,
Trinchieri, and MACDonald 1984).
The irregular galaxies were chosen primarily for their blue color and
in most cases display a disturbed morphology (Fabbiano, Feigelson, and
Zamorani 1982). 

\subsection{Early-Type Galaxy Pairs}

Eight early-type galaxy pairs are analyzed of which 
3 are detected with higher
than 3$\sigma$ significance. 
The sample of
normal early-type galaxies observed with {\it EINSTEIN} 
(Eskridge, Fabbiano, and Kim 1995a)
is shown in Figure 5 with the best fit line for comparison. 
A value of 50 km sec$^{-1}$ Mpc$^{-1}$ for
the Hubble Constant was used
in both samples and the pair luminosities were converted to
the energy band of the normal galaxies, 0.2 - 4 keV, using the
canonical spectrum described in section 2. 
These authors found that the sample is best characterized by two linear
relations and those with log l$_{x}$ $>$ 40.5 have a
best fit line with a steeper slope, $\simeq$2. This
attributed to
X-ray emission from low luminosity normal galaxies being
dominated by low mass X-ray binaries while the high l$_{x}$ 
have a dominant contribution from hot gas. All
of the pairs have l$_{x}$ $>$ 40.5; they are X-ray
bright and fall into the latter category. Compared to
the normal galaxies, the
early-type pairs appear
underluminous in X-rays for 
their l$_{b}$. 
We compare the observed l$_{x}$ for 
pairs with the sum of the predicted l$_{x}$ for each galaxy. The
predicted l$_{x}$ is calculated from the l$_{b}$ for the galaxies
individually, then added to get the predicted l$_{x}$ for the
pair. The relationship reported by Eskridge, Fabbiano, and
Kim (1995a) for the brightest ellipticals, l$_{x}$ $\sim$ l$_{b}^{2}$,
is used for the predicted values. Figure 6 shows that the observed
values systematically fall below the predicted values,
except in the case of the detected pair, CPG 564, for
which we report evidence of an intergalactic medium (section 4).
The mean predicted log l$_{x}$ is 42.10$\pm$0.19:
2$\sigma$, 95\% confidence, higher than the mean observed value of
the pairs, 41.35 $\pm$ 0.21. If CPG 564 is excluded, the
mean observed log l$_{x}$ decreases to 41.16 $\pm$0.12
and the
difference increases to 3$\sigma$.

The blue magnitudes from Karachentsev (1972) are isophotal and extend to
	the B=26 mag./sq. arcsec isophote. This is one magnitude fainter than the
	limiting value in RC2 which was used in the comparison sample of early
	type galaxie in Eskridge et al. (which is based on Fabbiano
	et al. (1992)). The NED data base was used to 
	obtain the RC3 magnititudes for the brightest galaxy in the 8 pairs.
	The magnitudes from the CPG and RC3 are both corrected for extinction
	in the Milky Way. The CPG pairs are additionally corrected for internal
	extinction based on galaxy type and inclination. The values are
	listed for comparison with the CPG magnitude given first: 
	CPG 18(14.78,14.60), CPG 90(15.39,15.7), CPG 99(12.91,12.61), CPG 175
	(13.21,12.12), CPG 320(14.49,14.9), CPG 367(14.52,13.64), CPG 564(14.63,
	14.32), CPG 574(13.98,13.63). The RC3 values for CPG 18, CPG 90, and
	CPG 320 are not corrected for extinction and CPG 90 and CPG 320
	are a little high. For all of the other pairs
	the RC3 magnitude is slightly lower; generally all of the pairs
	agree within the published RC3 error. The trend reported in this
	paper, that early- and mixed-type pairs have a lower l$_{x}$ for
	their l$_{b}$ (based on the CPG) can not be explained as a systematic
	overestimate of the l$_{b}$ from CPG relative the RC3, on which the
	camparison sample is based. On both figures 5 and 6, use of the RC3 magnitudes
	for the paired galaxies would generally shift the pairs to the right,
	making the underluminosity more evident.

The process of disk galaxy merger can give a blue
luminosity much higher than is typical of even the brightest isolated
galaxies (Caon et al. 1994). High l$_{b}$ galaxies occupy a
region of the fundamental plane consistent with mergers of systems
which are generally
less gaseous and more stellar (Bender et al. 1992). Thus,
those galaxies which have a high l$_{b}$ are associated
with interaction and in some cases, merger. The very high
l$_{b}$ early-type pairs are also characterized by a low X-ray luminosity
similar to NGC 4125 and NGC 3610. 
These galaxies also have low 0.1 - 2 keV X-ray
luminosities and have plumes of emission suggesting the later 
stages of a merger (Fabbiano and Schweizer 1995). 
The prototype for the pairs is the close elliptical pair, CPG 99, 
found to show evidence of tidal interaction in the form of
a U-shaped velocity dispersion (Borne and
Hoessel 1988; Bonfanti et al. 1995). 
In addition, the galaxy NGC 1587, which is part
of the pair was argued to be
a merger remnant on the basis of an angular momentum which could
not be attributed to the present interaction. The CPG 99 pair
is detected at $>$3$\sigma$ and has a
low l$_{x}$, $<$ 10$^{41}$ ergs sec$^{-1}$, for
it's l$_{b}$, $\sim$2$\times$10$^{44}$ ergs sec$^{-1}$. The early-type
pairs typically have low l$_{x}$ and may show evidence of being merger
remnants as well as interaction with their companion.
The NGC4782/4783 pair (Colina and Borne 1995) also shows a low integrated
X-ray luminosity, $\sim$2x10$^{41}$ ergs sec$^{-1}$, typical
of the underluminous pairs found here.
High resolution studies of the 4782/4783 system 
show multiple X-ray components 
resulting from the interaction and merger of the two galaxies
which includes tidal heating and shock heating from the
collision. 
This detailed study supports
	the hypothesis that in the underluminous pairs
	the X-ray emission is
	directly associated with the interaction/merger.
	The spatial analysis presented in section 4 contained CPG 99 and CPG 564.
	CPG 99 is underluminous by nearly a factor of 10, the
	least X-ray luminous detection. The spatial analysis 
	shows it has a very small radial extent of $\sim$ 47 kpc and small core
	radius when compared to similar parameters for groups.
	For example, the NGC 2300 group has a radial extent of $>$250 kpc.
	In contrast, the only pair with a luminosity over that
	predicted, CPG 564, has X-ray emission extending to $>$169 kpc,
	which is likely intergalactic. The gas temperature is
	1.03 +0.9/-0.26 keV, similar to that found for groups.
	If the underluminous early-type galaxy pairs such as CPG 99
	are merger remnants, then they must have formed out
	of spiral pairs with no intergalactic medium. Generally,
	these are the low velocity dispersion groups (Mulchaey et al. 1996).
Spectral fits performed on five detected paired galaxies 
gave significant temperature constraints with 
best fitting temperatures between 0.20 - 1.26 keV; in all
cases the 90\% range in temperature excludes a hard component
dominating the emission. Thus, the spectrum is dominated by
hot gas. 
{\it ASCA} and {\it ROSAT} 
spectra of normal elliptical galaxies show both a hard
component,
with l$_{x}$ correlated with l$_{b}$ (Matsushita et al. 1994) implying
a stellar origin, and a soft component which dominates
the emission of bright ellipticals 
(Fabbiano, Kim, and Trinchieri 1994; Kim et al. 1996). 
The soft component originates in hot gas with
the most X-ray luminous normal ellipticals being
consistent in their l$_{x}$-l$_{b}$ relationship with
a steady state cooling flow (Sarazin and Ashe 1989).

The gas dynamics in normal elliptical galaxies
fit a somewhat simple picture in which
the temperature of the hot gas component is proportional to the velocity 
dispersion of the galaxy but is
not in energy equipartition with the stellar component
implying that the gas is heated and bound by the dark matter dominated
gravitational potential (Davis and White 1996; Eskridge,
Fabbiano, and Kim 1995b). Studies of several pairs show that 
they are dynamically complex
with U-shaped velocity dispersion profiles, presumably from
tidal heating in the outer radii. Analogous to the stars, the
gas would also be tidally heated. Tidally triggered 
dynamical heating as well as that from star formation may be 
sufficient to overcome the binding gravitational potential
of the galaxy. This would account for the
paired galaxies being among the least luminous of those
normal early-type galaxies which are classified as X-ray luminous
and dominated by hot gas. Simple calculations by Mathews and Brighenti (1997)
suggest that a factor of $\sim$10 underluminous interacting early-type
galaxies must be recent merger/interactions, $\sim$10$^{9}$ years, due to
replenishment of the interstellar gas by stellar mass loss.
This would make 
gas deficiency an unlikely explanation for a large number
of low l$_{x}$/l$_{b}$
galaxies on statistical grounds. Figure 5 indicates that approximately
half of the pairs are underluminous
by a factor of 10 compared to the best fit linear regression
found for normal
early-type galaxies. However, paired galaxies are expected
to show a higher frequency of recent and ongoing interaction/merger
than single galaxies which may represent merger products so that
gas lost may not have been fully replenished by stellar mass loss.

Galaxies in early-type pairs show evidence of  
interaction and merger
based on very high blue luminosities and disturbed velocity 
dispersion profiles. They also tend to be deficient in X-rays
which, because of their soft spectra indicate a deficiency in
hot gas. This may have important
consequences for enrichment of the intergalactic medium with metals. One of the ways in 
which non-isolated galaxies
evolve, whether the environment is a pair, group, or cluster may be through
interactions with another galaxy. 
That interaction with a companion appears to destabilize gas within the pair system
so that it can not be bound by the galaxies. 
If the pairs were in a cluster or group environment,
the intergalactic medium
would be enriched during the merger process since the more massive
system would be able to bind the gas. These results would apply equally
well to a cluster of protogalaxies in which recent
observations of cluster abundances imply the enrichment must occur 
(Mushotzky and Loewenstein 1997).

\subsection{Mixed Type Galaxy Pairs}

There are 16 mixed morphology pairs of which
3 are detected at $>$3$\sigma$. 
Figure 7 shows observed l$_x$ versus 
predicted l$_x$ in the 0.2 - 4 keV band.  The
predicted values are the sum of the individual
values predicted for each galaxy using the
l$_{x}$ - l$_{b}$ appropriate for each galaxy
class.
There are
eight pairs: 2 detections and 6 upperlimits, with luminosities
below that predicted. 
Four (2 detections and 2 upperlimits) are significantly
underluminous. A spatial analysis of
the two detections, CPG 278 and CPG 353 
shows emission centered on both galaxies in the case
of 278 and on the elliptical in the case of CPG 353. The 90\%
confidence range in temperature for CPG 278 is 0.31 - 0.57 keV,
lower than all of the groups reported in Mulchaey et al. (1996).
The temperature of CPG 353 is 0.78 - 0.85 keV; however,
both
have X-ray emission about a factor of 10 below
that predicted for their l$_{b}$. They are underluminous
in the same sense that the early-type pairs are underluminous;
they are gas poor galactic systems and lack an additional
intergalactic component.

\section{Spatial Analysis}

Eight detections (excluding those dominated by AGN) are 
located in an appropriate
region of the detector and had sufficient counts 
for spatial analysis: 
CPG 234, CPG 239, CPG 278, and CPG 353 (mixed pairs), CPG 99 
and CPG 564
(early-type), CPG 125 and CPG 218 (spiral
pairs). The PSPC images were flatfielded, corrected
for vignetting, and background subtracted prior to further
analysis. 

Radial profiles were made using the X-ray peak as the center (Figures
8 - 15). These are used to obtain the extent of the X-ray
emission. 
X-ray contours are overlayed on the Digitized
Sky Survey image (Figures 16 - 21) to see
the relationship between the X-ray emission and the galaxies. 
The contours are smoothed with
a Gaussian of width 22.5 arcsec, chosen to minimize the
confusion between close galaxy pairs.

From the radial profiles, the peak surface brightness
and HWHM is measured as well as the
approximate radial extent.
For CPG 125, 218, and 278, the emission is resolved
into two distinct components: one centered on each galaxy.
There is an entry in Table 8 for each galaxy in these pairs. 
Intergalactic group X-ray emission is typically fit
by a $\beta$ of 0.5 using the radial profile, (1 + (r/r$_{c}$)$^{2}$)$^{-3\beta + 
0.5}$.
For this profile, r$_{c}$ is equal to the HWHM so that the
core radii measured by Mulchaey et al. (1996) using this profile
can be compared with the HWHM measured for the pairs.

The HWHM of CPG 125 and CPG 218, the spiral pairs is consistent
with a point source while for the other 6 pairs the emission is
extended. This comparison was made allowing for the broadening
PSF of the PSPC with radius.
The HWHM for the pairs with extended emission
are generally smaller than the core radii of groups.
The extent of the
gas is much smaller than groups, except in the case of CPG 564. CPG 564 is, in
fact, the most luminous early-type pair, 4.7$\times$10$^{42}$ ergs
sec$^{-1}$ and comparable to the brightest in the survey of
isolated galaxies. The extent of it's X-ray emission, $>$169 kpc, and
HWHM, $\sim$80 kpc, is comparable
to that from an intergalactic medium.
However, the X-ray profiles
of the other five indicate galactic emission.

The X-ray emission originating from hot gas
in early- and mixed-type pairs
show they are generally underluminous in X-rays 
compared to that expected from isolated
galaxies. 
There is also no spatial evidence that these
optically selected pairs have intergalactic gas
similar to the NGC2300 system.
If the early- and mixed-type galaxies formed
through mergers, they must have formed from spiral rich groups
with little or no intergalactic medium. These are generally groups with
a low velocity dispersion (Mulchaey et al. 1996).
 
\section{Conclusions}

We have completed the first X-ray survey of optically selected
galaxies in pairs using {\it ROSAT} pointed observations. 
An analysis of the hot gas content in
early- and mixed-type galaxy pairs 
indicates that they are generally characterized
by a lower l$_{x}$ for their l$_{b}$ compared to normal
galaxies. A spatial analysis 
indicates that the extent of 
the X-ray emission is much less than 
the intergalactic medium of a group and is
centered on a galaxy. This is
consistent with X-ray emission being galactic in origin.
The mean l$_{x}$
for early-type pairs is 2$\sigma$, 95\% confidence, 
lower than the mean predicted
if they were isolated galaxies implying
that they are gas poor. 
Mixed pairs show this same trend.
The X-ray brightest pair, CPG 564, is an exception to
this trend since a spatial and spectral analysis shows
its emission is 
likely intergalactic. These pairs show evidence of interaction
and possibly merger in other wavebands. If they have
formed through mergers, then the lack of any
intergalactic emission could be explained if they formed
out of spiral dominated, low velocity dispersion groups. These typically
have no intergalactic medium.

The spiral galaxy pair sample has a number of pairs
with significantly higher l$_{x}$/l$_{b}$ than the normal
galaxy sample, 
comparable to the brighter starburst 
galaxies and less luminous Seyfert
galaxies.  
 
X-ray studies of galaxies in pairs are of fundamental
importance to understanding the 
impact of interacting galaxies on galaxy evolution. 
Enrichment of the
intergalactic medium may also be tied to protogalactic
interactions and gas loss as reported here for early- and
mixed-type pairs. The relationship between the X-ray emission
of pairs
and galaxy groups of 3-4 members is also crucial to understanding
the dynamical history of the groups and any dependence on morphology.
Future detailed studies of individual
pairs of each galaxy type which provide spatially and spectrally
resolved X-ray components are necessary to 
discover the detailed interaction history of these interesting objects.

\acknowledgements
MH and SC gratefully acknowledge support for this project from the
National Science Foundation grant No. AST-9624716. We are thankful
to the referee, Trevor Ponman, for critical comments 
on the manuscript.

\clearpage

\begin{deluxetable}{ccccccccccc}
\footnotesize
\tablewidth{0pt}
\tablecaption{Data for Spiral Galaxy Pairs}
\tablehead{
\colhead{CPG}           & \colhead{Type}      &
\colhead{RA}          & \colhead{Dec}  &
\colhead{Sep}          & \colhead{Vel}    &
\colhead{L$_b$}  & \colhead{$Cnts \over sec$ $\times10^{-2}$ }  &
\colhead{$\sigma$} & \colhead{L$_{x}$} & \colhead{Log L$_{FIR}$} 
}
\startdata
3 	& 	S	& 	00 10 33 	& 	28 59 46 	& 	1.5	& 	7131	& 	3.41 	&  	1.49  		& 	1.55 	& 	$<$12.10	& 	44.12\nl
	& 	Sc	& 	00 10 26 	& 	28 59 14 	& 	-	&	7307 	& 	11.70 	& 	- 		& 	-	& 	- 	& 	-  \nl
7	&	S0? 	& 	00 21 13 	& 	30 28 26 	&	2.06	&	4909	&	2.73	&	0.27		&	0.95	&	$<$3.41 &	-\nl
	& 	SB?	& 	00 21 16 	& 	30 30 20 	&	-	&	4943	&2.69	&-	&-&	- & 43.29\nl
13 	& 	SA(s)ab	& 	00 36 52 	& 	23 59 29	& 	0.37	& 	4496 	& 	10.76	&  	0.28		& 	0.50	& 	$<$1.55 	& 	43.82 \nl
	& 	SAB pec	& 	00 36 52 	&	23 59 05 	& 	-	& 	4674	& 	2.06	& 	-	 	& 	- 	&	-  	& 	- \nl
31	&	S pec 	&	01 24 33 	&	03 48 03 	& 	0.65	& 	2276 	& 	5.87 	& 	0.77		& 	0.34	&	$<$0.26	& 	44.56\nl
	&	S	&	01 24 35 	&	03 47 22 	& 	-	&	2216 	&	2.30	& 	-		& 	-	& 	 - 	& 	-\nl
125	& 	pec	& 	07 09 12	& 	20 36 06	&  	2.57	& 	5051	& 	11.09	& 	2.40		& 	0.46	&	3.80 	& 	- \nl
	&	S pec	& 	07 09 17 	&	20 38 11 	& 	-	& 	5154	& 	30.10	&	-	 	& 	- 	& 	-	& 	44.92\nl
136	&	S? 	&	07 27 22 	&	19 38 22 	& 	1.00	&	9660 	& 	16.60	& 	0.47		& 	0.36	& 	$<$5.03	&	44.55 \nl
	& 	S?	&	07 27 24 	&	19 37 27 	&	- 	&	9821 	& 	14.80	& 	-		& 	-	& 	 -	&	- \nl
137 	& 	S 	& 	07 36 44 	& 	74 26 57 	&	0.46	&	3918	&	1.33	&	1.57		&	0.37	&	0.61   &	43.48\nl
	&	S	&	07 36 37	&	74 26 45	&	-	&3927	&1.04	&	-&	-&	-&-\nl
140	&	Sb 	&	07 44 07 	&	29 14 57 	& 	0.54	& 	4739	& 	5.03	& 	0.52		& 	0.45	& 	$<$1.53	&	43.69 \nl
	&	Sa 	&	07 44 09 	& 	29 14 50 	& 	-	&	4791 	& 	5.85	& 	-		& 	-	& 	- 	&	-\nl
161	&	SAB 	&	08 23 33 	& 	21 20 15	&	0.72  &	5318 	& 	3.03	& 	0.16		& 	0.13	&	$<$3.62 	&	43.05 \nl
	& 	SB	& 	08 23 34	& 	21 20 51	& 	-	& 	5291 	& 	2.11	&	-		&	-	&	- 	&	- \nl
163	&  	S?	&	08 29 15 	& 	55 31 21 	& 	0.88	&	7777	&	6.98	&	5.61		&	2.52	&	$<$22.80&	44.07 \nl
	& 	Sbc 	& 	08 29 21 	& 	55 29 29	&	-	&	7828	&	9.15	&	-		&	-	&	-	&	-\nl
\tablebreak
171 	& 	S?	&	08 46 01 	& 	12 47 09 	&	1.54	&	8833	&	15.08	&	-0.68		&	0.97	&	$<$11.2	& 	44.18 \nl
	& 	S? 	& 	08 45 55 	& 	12 46 57 	&	-	&	8788	&14.30	&-	&-&	- &-\nl
186 	& 	Sm	& 	09 05 48	& 	25 26 11	& 	0.78	& 	2627 	& 	0.55 	& 	0.54		& 	0.40	& 	$<$0.40	&	43.85 \nl
	& 	Sc	& 	09 05 48	& 	25 26 11	&	- 	&	2553 	& 	6.73	& 	-		& 	- 	& 	- 	&	43.96\nl
200 	& 	Sb 	& 	09 25 00 	& 	64 33 37 	&	2.01	&	5137	&	4.44	&	-0.91		&	0.69	&	$<$2.91& 	-\nl
	&	Sb	&	09 25 05	&	64 31 40	&	-	&	5485	&        3.19   & 	-		&	-	&	-	&	-\nl
204	& 	SB	& 	09 27 45 & 	12 17 15 & 	0.23	& 	8496 	& 	9.05 & 	1.14		&	0.59	& $<$5.20 & 44.19\nl
	&	Im	& 	09 27 44 &	12 17 14 & 	-	&	8546 	& 	7.25 & 	-	 	& 	- 	&	-  \nl
218 	&	I0 	& 	09 55 51 	& 	69 40 26 	& 	39.6	&	  95	&	0.68	&	116.80		&	2.97	&	0.60 	&-	44.61\nl
	& SA(s)ab	& 	09 55 34 	& 	69 04 11 	&	-	&388	&3.76	&	-&	-&	- &-\nl
246	&	Sa	&	10 43 30	&	12 05 14	&	2.76	&	7790	&	10.40	&	0.43		&	0.42	&	$<$3.75	&	- \nl
	&	Sc	&	10 43 39	&	12 03 36	&	-	&	7761	&	8.43	&	-		&	-	&	-	&	- \nl	
258	& 	pair? & 	10 59 05	&	72 38 00 	& 	0.87	&	8217	&	11.05	&	-0.093		&	0.38	&	$<$3.83 &	44.22 \nl
	&	Sa	&	10 58 57	&	72 38 33	&	-	&8179	&4.52	&-&	-&	-&	-\nl
347 	& 	SA(rs)bc	& 	12 36 32	& 	11 15 28	& 	1.3	& 	2112 	& 	8.54	& 	1.67		& 	0.48	& 	1.1 	&	44.37 \nl
 	& 	SA(rs)bc	& 	12 36 37	& 	11 14 20	& 	-	& 	2158 	& 	9.17	& 	-		& 	- 	& 	- 	&	-\nl
378	& 	Sb pec	& 	13 30 06	& 	-01 43 17	& 	3.72	& 	4121 	& 	10.97	& 	1.29 		&	0.38	& 	1.3 	&	 44.09\nl
	&	SAB(r)c	&	13 30 11	&	-01 39 51	&	-	&	4055	&	11.10    & 	-		&	-	&	-	&	-\nl
410 	& 	SAc	& 	14 03 23	& 	09 26 51	& 	1.57	& 	5638 	& 	11.13	& 	0.28 		& 	0.37	& 	$<$1.45	&	44.05\nl
 	& 	Sbc	& 	14 03 27	& 	09 27 59	& 	-	&	5637 	& 	8.71	& 	-		& 	- 	&	-	&	-\nl
\tablebreak
419   &	Sa pec 	&	14 13 20 &	-03 08 56&	3.82	& 	1954	& 	2.42	& 	32.44		& 	2.31	& 	7.80 	& -\nl
	& 	SAB(r)	& 	14 13 15 & 	-03 12 27 & 	-&	1782 	& 	2.11	&		-	& 	-	&  	-	& 43.77 \nl
422	& 	(R)SB(rs)0/A	& 	14 17 03	& 	36 34 17	& 	0.59	& 	3139 	& 	4.92	& 	0.47		& 	0.29	& 	$<$4.58	& 	43.31\nl
	& 	SA(s)bc	& 	14 17 05	& 	36 34 30	& 	-	&  	3145 	& 	1.68	& 	- 		& 		& 	- 	& 	-\nl
473 	& 	Sb 	& 	15 49 57 	& 	20 48 16 	& 	0.48	&	10781	& 	16.73	&	0.13		&	0.21	&	$<$3.63 	&	44.52 \nl
	& 	S	& 	15 49 59 	& 	20 48 31 	&	-	&	10803	&	12.30	&	-		&	-	&	-	& 	-\nl
534	& 	Sb pec 	& 	18 12 56	& 	68 21 45 	&	0.73	&	6495	&	9.10	&	0.44		&	0.83	&	$<$0.54 & 	44.85\nl
	& 	Sa 	& 	18 12 60 	& 	68 21 16 	&	-	&6716	&11.30	&-	&-&	- &-	\nl
557 	& 	SBa 	& 	21 28 58 	& 	11 21 58 	& 	0.67	&	8646	&	9.37	&	0.24		&	0.29	&	$<$3.23	&	44.05\nl
	& 	SB	&	21 28 59 	& 	11 22 59 	&	-	&	8665 	&	8.43 	&	-		&	-	&	-	&	-\nl
566 	& 	SB? 	& 	22 19 28 	& 	29 23 45 	&	0.84	&	4994	&	2.98	&	0.043		&	0.58	&	$<$2.04 &	44.63 \nl
	&	S?	&	22 19 30	&	29 23 14	&	-	&	4768	&	2.93	&	-		&	-	&	-	&	-\nl
\enddata
\end{deluxetable}
\clearpage

\begin{deluxetable}{ccccccccccc}
\footnotesize
\tablewidth{0pt}
\tablecaption{Data for Early-type Galaxy Pairs}
\tablehead{
\colhead{CPG}           & \colhead{Type}      &
\colhead{RA}          & \colhead{Dec}  &
\colhead{Sep}          & \colhead{Vel}    &
\colhead{L$_b$}  & \colhead{$Cnts \over sec$$\times10^{-2}$ }  &
\colhead{$\sigma$} & \colhead{L$_{x}$} 
}
\startdata
18 & 	E & 	00 48 31 & 	01 21 17 & 	0.34 	& 	18648	& 	36.30 &  	0.82  	& 	0.34 & 	$<$14.9\nl
   & 	E & 	00 48 30 & 	01 21 12 & 	-	&	19003	& 	54.40 & 	- 		& 	-	 & 	-  \nl
90 & 	E& 	03 25 04 & 	02 53 46 & 	1.07	& 	9487	& 	7.73 &  	0.18		& 	0.16	 & $<$1.65 \nl
& 	E& 	03 25 01 & 	02 54 27 & 	-	& 	9159	& 	7.08 & 	-	 	& 	- 	 &	-  \nl
99&	E pec &	04 30 40 &	00 39 42 & 	0.97	& 	3800 	& 	12.20 & 	1.25		& 	0.359	 &	0.94\nl
&	E pec &	04 30 44 &	00 39 52 & 	-	&	3495 	&	 4.67& 	-		& 	-	& 	 - \nl
175& 	E1-2& 	08 49 22 & 	19 04 29 & 	 0.56  & 	4206	& 	11.30& 	2.05		& 	0.64	&	3.35 \nl
&	E0 pec & 	08 49 24 &  	19 04 26 & 	-	& 	3850	& 	 3.67&	-	 	& 	-	& 	- \nl
320 & E& 	12 05 41 & 	01 35 37 & 	1.10	& 	5959 	& 	5.45 & 	0.064		& 	0.24	& $<$0.98 \nl
 & 	SB(s)0& 	12 05 41 & 	01 34 29 & 	-	&	5550 	& 	4.95 & 	-		& 	- 	&	-\nl
367 & S0?& 	13 14 52 & 	17 13 36 & 	1.28  & 	7087 	& 	9.62 & 	-0.80		 & 	2.73	 & $<$16.60 \nl
& 	E& 	13 14 57 & 	17 13 33 & 	-	&  	7012 	& 	6.11 & 	- 		& 	-	 & 	- \nl
564&	SA0- &	22 14 47 & 	13 50 25 &	0.59 	& 	8212	& 	11.70& 	6.07		 & 	0.76 & 	46.5 \nl
& 	SA0-&	22 14 45 & 	13 50 45 &	- 	&	7964 	& 	11.70& 	-		 & 	- 	& 	- \nl
574 & S0 & 	22 51 00 & 	31 22 28 & 	0.57	&	6778	& 	14.5 &	0.48		&	1.32	&$<$7.87\nl
    & E	&  	22 51 04 & 	31 22 28 &	-	&	7216	&	3.90 &	-		&	-	&	- \nl
\enddata
\end{deluxetable}
\clearpage
\begin{deluxetable}{ccccccccccc}
\footnotesize
\tablewidth{0pt}
\tablecaption{Data for Mixed Galaxy Pairs}
\tablehead{
\colhead{CPG}           & \colhead{Type}      &
\colhead{RA}          & \colhead{Dec}  &
\colhead{Sep}          & \colhead{Vel}    &
\colhead{L$_b$}  & \colhead{$Cnts \over sec$ $\times10^{-2}$ }  &
\colhead{$\sigma$} & \colhead{L$_{x}$} & \colhead{Log (l$_{FIR}$)}
}
\startdata
83 	& 	RingA 	& 	02 55 10 & 	-00 10 41 & 	0.71 	& 	8514	& 	37.20 	&  	0.17 		& 	0.26 	& $<$2.33 	& 45.04\nl
	& 	RingB 	& 	02 55 12 & 	-00 11 00 & 	-	&	8714	& 	23.90	& 	- 		& 	-	 & 	-  	&-\nl
116 	& 	Sc	& 	07 02 31 & 	86 34 47 & 	1.57	& 	5005	& 	2.00 	&  	0.22		& 	0.34	 & $<$1.04 	& 43.75\nl
	& 	E	& 	07 03 21 & 	86 33 29 & 	-	& 	4951	& 	4.16 	& 	-	 	& 	- 	 &	-  	& -\nl
144 	& 	E	& 	07 48 13 & 	28 13 30 & 	0.54	& 	8107 	& 	9.99 	& 0.64 	& 0.54& $<$4.36& - \nl
	& 	Sa	& 	07 48 11 & 	28 13 50 & 	-	&	8210 	& 	4.69	 & -	& 	- &-	& 44.66 \nl
234	& 	E2	& 	10 23 27 & 	19 53 50 &  	2.35  & 1168	& 1.00  	& 26.29	& 	0.45&	11.& - \nl
	&	SAB(s) pec 	& 	10 23 31 &  19 51 54 & 	-	& 1063	& 2.09 	&-& -& 	- & 43.39\nl
238	&	E	&	10 34 30 &	39 36 54 & 	2.34	&	12862 & 	13.70 & 	331.70		& 	3.17	& 	867. & -\nl
	& 	compact	&	10 34 39 &	39 38 29 &	- 	&	12951	& 	13.90& 	-		& 	-	& 	 - & -\nl
239	&	E? 	&	10 36 21 &	58 37 11 & 	3.92	& 	8216 	& 	15.30 	& 	1.82		& 	0.92	 &	$<$7.65 	& -\nl
	&	SABbc 	&	10 36 26 &	58 33 22 & 	-	&	8252 	&	 9.98	& 	-		& 	-	& 	 - 	& -\nl
243	&	S0 	&	10 40 45 &	39 04 25 & 	1.08	&	9015  & 	3.66 	& 	0.19		& 	0.19	& $<$1.91	& -\nl
	& 	S 	&	10 40 40 &	39 03 57 &	- 	&	8914	& 	11.50 	& 	-		& 	-	& 	 -	& 44.29\nl
260	&	E?	&	10 59 59 &	50 03 24 & 	2.72	& 	7235	& 	9.66	& 	0.44		& 	0.39	& $<$2.54 	& -\nl
	&	S? 	&	10 59 53 & 	50 00 53 & 	-	&	7434 	& 	13.90 	& 		-	& 	-	& 	- 	& -\nl
278	&	SA(s) &	11 16 55 &	18 03 04 & 	5.96	& 	841	& 	7.44& 	8.52		& 	0.71	& 	0.45 & - \nl
	&	E &	11 16 59 & 	18 08 54 & 	-	&	1118 	& 	3.40& 		-		& 	-	& 	- \nl
339 	&	SB0?	& 	12 28 13 & 	13 53 56 & 	1.57  & 	7041 	& 	8.42 	& 	-1.94		 & 	0.71	 &  $<$4.44 	& -\nl
	&	Scd	& 	12 28 07 & 	13 54 43 & 	-	&  	7269 	& 	10.80 	& 	- 		& 	-	 & 	- 	& 43.72\nl
\tablebreak
345 	& 	S0/a	& 	12 32 48 & 	63 56 22 & 	4.02  & 	2591 	& 	5.46 	& 	0.055		 & 	0.32	 & $<$ 0.32	& - \nl
	& 	SB(s)dm	& 	12 32 34 & 	63 52 38 & 	-	&  	3073 	& 	2.02 	& 	- & 	- & 	-& 43.00\nl
353	&	E2 	&	12 43 40 & 	11 33 07 &	2.57  &	1170 	& 	20.30	& 	43.44		& 	1.00	&	3.1 	& -\nl
	& 	SAB(rs)c	& 	12 43 33 & 	11 34 57 & 	-	& 	1382	& 	3.83	&		-	&	-	&	-	& 43.50 \nl 
402	&	S0 	&	13 55 59 & 	17 29 57 &	0.73 	& 	6202	& 	2.10	&	 0.40		 & 	0.22 & 	$<$1.11	 & - \nl
	& 	S	&	13 56 00 & 	17 30 40 &	- 	&	6528 	& 	4.69	& 	-		 & 	- 	& 	- 	& 43.89\nl
494 	& 	E 	& 	16 17 36 & 	46 04 57 & 	1.15	&	6102	& 	2.25 	&	-0.68		&	1.13	&	$<$5.00& - \nl
    	& 	Sab	&   	16 17 31 & 	46 05 30 &	-	&	5919	&	4.73	&	-&-&- & 44.21\nl
508 	& 	E0	& 	17 19 14 & 	48 58 50 & 	3.79	& 	7488 	& 	5.69 	& 	0.65		 & 	0.32	 & 	$<$2.14	& -\nl
 	& 	Sb	& 	17 19 21 & 	49 02 50 & 	-	& 	7420 	& 	8.30  & 	-		& 	- 	& 	-  	& 43.92\nl
510	&	 E	& 	17 22 39 & 	30 52 11 & 	0.77	& 	13694	& 	20.50 & 	33.80		& 	2.69	& 	315. & -\nl
	& 	SAB(r)a	& 	17 22 40 & 	30 52 52 &	- 	&	13215	& 	20.70& 	-		 & 	-	 & 	& - \nl
530 	& 	E	& 	18 10 58 & 	31 06 56 & 	9.27	&	7227	& 	14.50 &	-0.50		&	0.72	&	$<$4.60 & -\nl
    	& 	Scd	&   	18 10 27 & 	31 00 11 &	-	&	7238	&	8.19	&	-&	-&	-& 43.91\nl
548	& 	E+pec	& 	20 47 24 & 	00 18 02 & 	1.76	& 	4032 	& 	5.99 	& 	0.95		&	0.47 	& $<$1.03 	& -\nl
	&	SAB(r)ab	& 	20 47 19 &	00 19 11 & 	-	&	4419 	& 	15.30	 &	- & 	- 	&	- 	& 43.53 \nl
\enddata
\end{deluxetable}
\clearpage

\begin{deluxetable}{ccccccccc}
\tablewidth{0pt}
\tablecaption{Supplemental Data for All Pairs}
\tablehead{
\colhead{(1)}           & \colhead{(2)}      & 
\colhead{(3)}  & \colhead{(4)}        & \colhead{(5)}  & \colhead{(6)} &
\colhead{(7)}          & \colhead{(8)}   &
\colhead{(9)} 
}
\startdata
3 	& wp200645	& 3.5 &	51.0	&2717 	& 0.32	& 	1.67	&	4.7	&  N27, U96, C0007.9+2844 \nl
	& 		& 	&  		&		& 		&		&		& U95, C0007.9+2843	 \nl
7	&rp201507n00& 4.4 &	47.0	&	4849	&0.10		&	1.77	&	5.8			&C0018.6+3011		\nl
	& 		& 	& 		& 		&		&		&				&C0018.7+3014		 \nl
13 	& rp800464n00& 3.8 &	3.3	&	7709	& 0.19 	&	1.04	& 	3.3 	& N169, U365,C0034.2+2343	\nl
	& 		&	&		& 		&		& 		& 		&N169A,C0034.2+2343		\nl
18	& rp700377n00& 2.9 & 	38.13	&	10447	&0.79 	&	1.41	&	2.6	&I1559,U496,C0045.9+0105	\nl
	&		&	&		& 		&		&		&		&U496,C0045.9+0105		\nl
31	& wp701047n00& 4.0 &	0.50	&	13894	& 0.75 &	1.03	& 	2.9	&N520,U966,C0122.1+0333 		\nl
	&		&		&		& &		&		&&N520,U966,C0122.1+0333   	\nl
83	& rp701403n00& 2.3 &	21.8	&	11863	&0.22 &	1.63	&	5.5	&N1143,U2388,C0252.6-0023	\nl		
	&		&		&		& &		&		&&N1144,U2389,C0252.6-0023	\nl
90	& rp700099	& 3.0 &	33.79	&	25727	& 0.39 &	1.71	&	8.8	&C0322.4+0243		\nl
	&		&		&		&	&	&		&&C0322.4+0244		\nl
99	& rp800471n00  &	3.3 & 15.28	 &	10640	&1.16 &	1.17	&	0.5	&N1587,U3063,C0428.1+0033\nl
	&		&		&		&	&	&		&&N1588,U3064,C0428.2+0033\nl
116	&rp900512n00  & 3.6 &	55.20	&	9034	&0.22 &	2.21	&	6.6	&U03528A,C0642.0+8640	\nl
	&		&		&		& &		&		&&U3536A,C0643.0+8638	\nl
125	& wp701047n00& 3.4 &	1.46	&	9982	&1.74 &	1.03	&	5.4	&N2341,U3708,C0706.3+2040 	\nl
	&		&		&		&	&	&		&&N2342,U3709,C0706.4+2043 	\nl
136	& wp500113	& 2.6 &	31.13	&	4487	&	0.44 &1.38	&	6.9	&C0724.2+1944		\nl
	&		&		&		& &		&		&&C0724.5+1943			\nl
\tablebreak
137	& rp800230n00& 3.3 &	24.10	&	8782	&1.40 &	1.89	&	3.6	&U3906,C0730.4+7434 \nl
	&		&		&		&&		&		&&U3906,C0730.4+7437	\nl
140	& wp200896	& 3.0 &	24.71	&	8046	&0.38 &	1.48	&	4.7	&U3995,C0741.0+2921	\nl
	&		&		&		&&		&		&&U3995,C0741.0+2921	\nl
144	& wp200175	& 3.4 &	40.33	&	8837	&0.40 &	1.48	&	3.4	&U4030,C0745.1+2820    \nl
	&		&		&		&&		&		&&U4030,C0745.1+2820		\nl
161	& rp600542n00 & 2.8 &	44.29	&	2714	&0.39 &	1.78	&	4.3	&I2338,U4383,C0820.7+2130	\nl
	&		&		&		& &		&		&&I2339,U4383,C0820.7+2130	\nl
163	& rp701397n00& 2.9 & 43.54	&	1713	&0.74 &	1.77	&	4.5	&C0825.4+5540	\nl
	&		&		&		&&		&		&&U4427,C0825.3+5540 	\nl
171	& rp300203n00 & 3.3 &	23.29	&	4908	&-0.24&	1.85	&	3.8	&C0843.3+1259	\nl
	&		&		&		&&		&		&&C0843.1+1258	\nl
175	& wp700541	& 3.8 &	39.39	&	6062	&1.07&	1.58	&	0.24	&N2672,U4619,C0846.5+1916	\nl
	&		&		&		&&		&		&&N2673,U4620,C0846.5+1916	\nl
186	& wp800602n00& 3.5 &	12.97	&	7535	&0.45 &	1.09	&	3.5	&N2750,U4769,C0902.8+2538	\nl	
	&		&		&		&&		&		&&N2750,U4769,C0902.8+2538	\nl
200	& rp800466n00& 3.6 &	33.74	&	6624	&-1.31&	1.42	&	3.6	&C0921.0+6447		\nl
	&		&		&		&	&	&		&&C0921.1+6445	\nl
204	& wp701039n00& 2.6 &	36.43	&	5247	&0.65 &	1.4	&	3.5	&U5044,C0925.0+1230	\nl
	&		&		&		&&		&		&&U5044,C0925.0+1230		\nl
\tablebreak
218	& wp600382n00& 39.0 &	19.00	&	28350	&13.1  &1.37	&	20.0	&M82,N3034	\nl
      &                 &           &           &      &          & &          &U5322,C0951.7+6955 \nl
	&		&		&		& &		&		&&M81,N3031\nl
      &           &           &           &  &        &           &&U5318,C0951.4+6918  \nl
234	& rp200076	& 10.8 & 53.00	&	26522	& 19.30&	2.7	&	3.1	&N3226,U5617,C1020.7+2008	\nl
	&		&		&		&&		&		&&N3227,U5620,C1020.8+2006	\nl
238	& wp700551	& 4.3 & 1.27	&	4647	&34.88&	1.0	&	1.8	&C1031.6+3953	\nl
	&		&		&		&	&	&		&&C1031.7+3954	\nl
239	& rp300204n00& 5.6 &	23.11	&	3395	&0.65 &	1.5	&	6.6	&N3286,C1033.1+5853	\nl
	&		&		&		&&		&		&&N3288,U5752,C1033.2+5849	\nl
243	& wp800410	& 3.1 & 52.75	&	10748	&0.32 &	2.2	&	1.6	&C1037.8+3920	\nl
	&		&		&		&&		&		&&C1037.8+3920		\nl
246	& rp700384n00& 3.8 &	12.04	&	9735	&	0.34 &1.12	&	2.8	&C1040.9+1222		\nl
	&		&		&		&	&	&		&&C1041.0+1221		\nl
258	&rp700872n00& 3.4 &	35.07	&	13081	&	-0.08 &1.43	&	3.7	&M159,C1055.4+7253		\nl
	&		&		&		&&		&		&&C1055.3+7254		\nl
260	& wp300137	& 3.4 & 35.60	&	9395	&0.38 &	1.39	&	1.3	&C1057.0+5019		\nl
	&		&		&		&	&	&		&&C1056.9+5017		\nl
278	& rp600263n00& 9.5 &	3.00	&	23865	&4.02 &	1.05	&	2.4	&N3607,U6297,C1114.3+1819	\nl
	&		&		&		&&		&		&&N3608,U6299,C1114.4+1825	\nl
\tablebreak
320	& rp600519n00& 2.5 &	31.17	&	21288	&	0.09 &1.54	&	1.9	&C1203.2+0153	\nl
	&		&		&		&	&	&		&&C1203.2+0152		\nl
339	& rp700346n00& 3.2 &	21.77	&	7069	&-2.74&	1.78	&	2.8	&N4447,C1225.7+1410	\nl
	&		&		&		&&		&		&&N4446,U7586,C1225.6+1441	\nl
345	& rp800522n00& 4.7 &	51.94	&	16836	&0.06 &	1.85	&	1.9	&N4521,U7706,C1230.6+6413	\nl
	&		&		&		&&		&		&&N4512,U7700,C1230.4+6410	\nl
347	& rp700056n00& 5.0 &	38.24	&	8990	&1.17 &	1.46	&	0.10	&N4567,U7777,C1234.0+1132	\nl
	&		&		&		&&		&		&&N4568,U7776,G1234.0+1131	\nl
353	& rp600017n00& 7.5 &	1.82	&	13666	&14.44 &	1.02	&	2.2	&N4649,U7898,C1241.1+1150	\nl
	&		&		&		&&		&		&&N4647,U7896,C1241.0+1152	\nl
367	& wp201372	& 3.3 &	29.75	&	1710	&-2.93 &	1.82	&	1.8	&I858,U8321,C1312.4+1729	\nl
	&		&		&		&	&	&		&&I859,U8320,C1312.5+1729	\nl
378	& wp800553n00& 4.3 &	14.87	&	13148	&1.13&	1.05	&	3.0	&N5183,U8485,C1317.5-0128	\nl
	&		&		&		&&		&		&&N5184,U8487,C1327.6-0125	\nl
402	& rp700392n00& 3.3 &	52.00	&	11095	&0.59 &	2.00	&	2.0	&I960A,U8849,C1353.6+1745	\nl
	&		&		&		&&		&		&&I960B,U8849,C1353.6+1745\nl
410	& rp700345n00& 3.1 &	15.21	&	7298	&0.24 &	1.11	&	1.9	&N5434,U8965,C1400.9+0941	\nl
	&		&		&		&&		&		&&U8967,C1401.0+0943	\nl
419	& rp700332n00& 6.8 &	3.05	&	4251	&4.69 &	1.03	&	326.	&M1376,N5506,C1410.7-0259	\nl
	&		&		&		&&		&		&&N5507,C1410.8-0255	\nl
\tablebreak
422	& wp600462	& 3.4 &	27.63	&	14246	&0.54 &	1.34	&	1.1	&N5544,U9142,C1415.0+3648	\nl
	&		&		&		&&		&		&&N5545,U143,C1415.0+3648	\nl
473	& rp701373n00& 2.6 &	37.54	&	15434	&0.21 &	1.48	&	4.1	&U10052,C1547.8+2058	\nl
	&		&		&		&	&	&		&&U10052,C1547.8+2058	\nl
494	& rp200993	& 2.9 & 26.44	&	3905	&-.20&	1.34	&	1.1	&U10325,C1616.0+4613	\nl
	&		&		&		&&		&		&&U10325,C1616.0+4613	\nl
508	& wp701080n00& 4.1 &	56.36	&	16066	&0.69 &	3.05	&	214.	&U10814,C1717.9+4901	\nl
	&		&		&		&&		&		&&U10814,C1718.0+4905	\nl
510	& wp701086	&2.9 & 0.79	&	1922	&4.20 &	1.03	&	1.5	&C1720.7+3056		\nl
	&		&		&		&&		&		&&C1720.7+3055	\nl
530	& wp400043	& 7.4 & 24.96	&	15144	&-0.23&	1.44	&	6.0	&U1135,C1808.6+3059		\nl
	&		&		&		&&		&		&&N6575,U11138,C1809.0+3105	\nl
534	& rp800498n00& 3.6 &	50.93	&	5412	&0.18&	1.93	&	5.3	&N6621,U11175,C1813.2+6821		\nl
	&		&		&		&&		&		&&N6622,U11175,C1813.2+6820		\nl
548	& wp500308	& 3.4 & 30.66	&	8690	&0.67 &	1.33	&	5.7	&N6964,U11629,C2044.9+0008		\nl
	&		&		&		&&		&		&&N6962,U11628,C2044.8+0008		\nl
557	& rp400081	& 3.1 & 50.14	&	8786	&0.28 &1.95	&	5.6	&U11751,C2126.6+1110	\nl
	&		&		&		&&		&		&&U11751,C2126.6+1110	\nl
564	& wp700525	& 5.4 & 48.01	&	7409	&2.67&	1.56	&	5.5	&N7237,U11958,C2122.3+1335		\nl
	&		&		&		&&		&		&&N7236,U11958,C2122.3+1335		\nl
\tablebreak
566	& wp701046n00& 3.7 & 	0.00	&	9192	&0.02&	1.02	&	6.5	&N7253A,U11984,C2217.2+2098	\nl
	&		&		&		&&		&		&&N7253B,U11985,C2217.2+2098		\nl
574	& rp200986n00& 2.9 &	25.04	&	3491	&0.12&	1.21	&	6.9	&U12214,C2248.6+3106	\nl		
	&		&		&		&&		&		&&C2248.7+3106	\nl
\enddata
\tablenotetext{}{Column (1) - CPG Number}
\tablenotetext{}{Column (2) - PSPC Sequence Number}
\tablenotetext{}{Column (3) - Radius of circle}
\tablenotetext{}{Column (4) - Angle from center of image in arcminutes}
\tablenotetext{}{Column (5) - Exposure time in seconds}
\tablenotetext{}{Column (6) - Signal-Noise/3 in PSPC counts/second}
\tablenotetext{}{Column (7) - Exposure map correction}
\tablenotetext{}{Column (8) - Column density in 10$^{20}$ cm$^{-2}$}
\tablenotetext{}{Column (9) - Common galaxy names: (N)GC, (U)GC, (I)C, (C)GCG}
\end{deluxetable}
\clearpage

\begin{deluxetable}{cccccccccc}
\footnotesize
\tablewidth{0pt}
\tablecaption{X-ray Spectral Results}
\tablehead{
\colhead{CPG}   & \colhead{No. PHA Bins} & \colhead{$\chi^{2}_{dof}$}      &
\colhead{Best Fit Temperature (keV)}          & \colhead{90\% Confidence Range (keV)}  
}
\startdata
99	& 6	  	&	0.36	&0.99		&0.44 - 64    \nl
125	& 11		&	0.16	&1.09		&   0.37 - 64		\nl
137	& 6		& 0.37	& 0.83	&0.06 - 47.    \nl
175	& 6		&	0.04	& 0.68	& 0.27 - 64	\nl
218	& 151		& 0.48	& 0.85	& 0.70 - 1.14	\nl
234	& 162		&	0.85	& 2.09	& 1.67 - 2.73			\nl
278	& 84		&	33.9	& 0.40	& 0.31 - 0.57	\nl
347	& 10		&	0.33 &	1.71	&0.31 - 64	\nl
353	& 117		&	0.60	&0.82		&0.78 - 0.85		\nl
378	& 8 		&	0.11	&1.07	& 0.10 - 64	\nl
564	& 20		&	0.34	&	1.03	&0.77 - 1.93	\nl
\enddata
\tablenotetext{}{Column (5) - the value of 64 keV is the hard limit imposed
by the spectral fitting routine}
\end{deluxetable}
\clearpage

\begin{deluxetable}{cccccccccc}
\footnotesize
\tablewidth{0pt}
\tablecaption{Correlation Test Results: All Pairs}
\tablehead{
\colhead{DV}    & \colhead{IV}  & \colhead{DET/TOT}      & 
\colhead{GKT}          & \colhead{Slope}  & \colhead{Error} &
\colhead{IC}   & \colhead{Error}   
}
\startdata
log (l$_{x}$) (0.25 - 2 keV) 	& log (l$_{b}$) 	& 11/49 & 0.81 & 0.25	& 0.49 & 29.78 \nl
log (l$_{x}$)			& log (l$_{FIR}$)	& 7/36  & 0.40 & 	-	& -	& - & - \nl
log (l$_{x}$) 			& log ($\Delta$r)	& 11/49 & 0.69 & -0.51  & 0.65	&  40.20 \nl
log (l$_{x}$) 			& log ($\Delta$v)	& 11/49 & 0.52 &  -	& -	&  -\nl
\enddata
\end{deluxetable}
\clearpage

\begin{deluxetable}{cccccc}
\footnotesize
\tablewidth{0pt}
\tablecaption{Mean Values of Sample Quantities}
\tablehead{
\colhead{Sample}    & \colhead{Quantity}        & \colhead{Mean}      &
\colhead{Error}       
}
\startdata
spiral pairs & log (l$_{x}$)\tablenotemark{a} & 40.82&	0.11  \nl
normal spirals &log (l$_{x}$) \tablenotemark{b}& 39.87  & 0.11 \nl
early-type pairs & log (l$_{x}$)\tablenotemark{c} & 41.35 &	0.21  \nl
normal early-type: & log (l$_{x}$)\tablenotemark{d}& 42.00 & 0.13 \nl
(log l$_{x}$ $>$ 43.70) & & \nl
mixed pairs & log (obs. l$_{x}$) \tablenotemark{c}& 40.87 &	0.11 \nl
all pairs& log (l$_{x}$) \tablenotemark{e}& 40.86 &	0.09\nl
compact groups & log (l$_{x}$) \tablenotemark{e}& 41.59 &	0.11\nl
\enddata
\tablenotetext{a}{{\it ROSAT} PSPC data: extrapolated to 0.5 - 3 keV}
\tablenotetext{b}{{\it EINSTEIN} IPC data: 0.5 - 3 keV}
\tablenotetext{c}{{\it ROSAT} PSPC data: extrapolated 0.2 - 4 keV}
\tablenotetext{d}{{\it EINSTEIN} IPC data: 0.2 - 4 keV}
\tablenotetext{e}{{\it ROSAT} PSPC data: 0.25 - 2 keV}
\end{deluxetable}
\clearpage

\begin{deluxetable}{cccccc}
\footnotesize
\tablewidth{0pt}
\tablecaption{X-ray Spatial Results}
\tablehead{
\colhead{CPG}   & \colhead{Peak Surface Brightness} & \colhead{HWHM(kpc)}      &
\colhead{Radial Extent(kpc)} 
}
\startdata
99	& 3.6		&	$\sim$14	& $\sim$47  \nl
125	& 24.7 	&	5	&	17  \nl
	& 15.1	&	10	&	25  \nl
218	& 14.5	&	1	&	2	\nl
	& 53.4	& 1	&		2		\nl
234	& 7.9	&	15 	& 40 \nl
239 	& 13.0	&	32 &	     74 \nl
278	& 12.6	&	2& 7 	\nl
	& 9.2	&	3& 9	\nl
353	& 471.	&	2& 6	\nl
564	& 1.4  	& 80 &	$\sim$169 \nl
\enddata
\tablenotetext{}{Column (2): 10$^{-3}$ counts sec$^{-1}$ cm$^{-2}$}
\end{deluxetable}
\clearpage

\figcaption[]{Log (l$_{x}$) vs Log (l$_{b}$) is shown for
all galaxy pairs. The filled hexagons are detected spiral pairs, the crosses are
detected early-type galaxy pairs, the filled triangles are detected mixed pairs, and the open
squares are upperlimits for all pair types. Detected pairs with
a Seyfert galaxy are represented by an open star. They are shown
here but not included in correlations. The X-ray luminosities
are given in the 0.25 - 2.0 keV energy band.}\label{fig1}

\figcaption[]{Log (observed l$_{x}$) vs Log (l$_{FIR}$) is shown for 
detections (filled hexagons) 
and upperlimits (open squares). CPG 419 is detected
as has a spiral galaxy. It was shown as a star
but not included in the correlations.}\label{fig2}

\figcaption[]{Log (observed l$_{x}$) vs Log ($\Delta$r) is shown for 
detections (filled hexagons) 
and upperlimits (open squares). Detected pairs
with Seyfert galaxies are shown as stars. They are
not included in the correlations.}\label{fig3}

\figcaption[]{Log (l$_{x}$) vs Log (l$_{b}$) is shown for detected spiral 
galaxy pairs (filled hexagons), normal spiral galaxies (crosses), and 
spiral pair upperlimits (open hexagons). Pair luminosities are converted
to the energy band of the normal galaxies,
0.5 - 3 keV. CPG 419 is detected and has a Seyfert
galaxy; it is shown by a star.}\label{fig4}

\figcaption[]{Log (l$_{x}$) vs Log (l$_{b}$) is shown for detected early-type
galaxy pairs (filled hexagons) and detected normal early-type galaxies (filled triangles).
The open hexagons are upperlimits for pairs while the
open triangles are normal early-type galaxy upperlimits. The pair
luminosities are converted to the energy band 
of the normal galaxies, 0.2 - 4 keV.}\label{fig5}

\figcaption[]{Log (observed l$_{x}$) vs Log (predicted 
l$_{x}$) is shown for early-type
galaxy pair detections (filled hexagons) and 
upperlimits (open hexagons). Luminosities are converted to
0.2 - 4 keV band.}\label{fig6}

\figcaption[]{Log (observed l$_{x}$) vs Log (predicted 
l$_{x}$) is shown for mixed
galaxy pair detections (filled hexagons) and 
upperlimits (open hexagons). Luminosities are converted to
0.2 - 4 keV band. CPG 238 and CPG 510 are detected and have
Seyfert galaxies; they are shown by a star. }\label{fig7}

\figcaption[]{Radial profile of CPG 99.}\label{fig8}

\figcaption[]{Radial profile of CPG 125 on the brightest
galaxy.}\label{fig9}

\figcaption[]{Radial profile of CPG 218 on the brightest
galaxy.}\label{fig10}

\figcaption[]{Radial profile of CPG 234.}\label{fig11}

\figcaption[]{Radial profile of CPG 239.}\label{fig12}

\figcaption[]{Radial profile centered on the brightest galaxy in 
CPG 278.}\label{fig13}

\figcaption[]{Radial profile of CPG 353.}\label{fig14}

\figcaption[]{Radial profile of CPG 564.}\label{fig15}

\figcaption[]{Contour map of CPG 99 is shown for image smoothed with
a Gaussian of width 22.5 arcsec. 
Contour levels are 2 and 3$\sigma$.}\label{fig16}

\figcaption[]{Contour map of CPG 234 is shown for image smoothed with
a Gaussian of width 22.5 arcsec. Contour levels are 3, 5, 
and 10$\sigma$.}\label{fig17}

\figcaption[]{Contour map of CPG 239 is shown for image smoothed with
a Gaussian of width 22.5 arcsec. Contour levels are 1, 2, 
and 3$\sigma$.}\label{fig18}

\figcaption[]{Contour map of CPG 278 is shown for image smoothed with
a Gaussian of width 22.5 arcsec. Contour levels are 3, 5, and 
10$\sigma$.}\label{fig19}

\figcaption[]{Contour map of CPG 353 is shown for image smoothed with
a Gaussian of width 22.5 arcsec. Contour levels are 3, 5, 10, 
20, 30, 40$\sigma$.}\label{fig20}

\figcaption[]{Contour map of CPG 564 is shown for image smoothed with
a Gaussian of width 22.5 arcsec. Contour levels are 1 and 
2$\sigma$.}\label{fig21}
\clearpage
\plotone{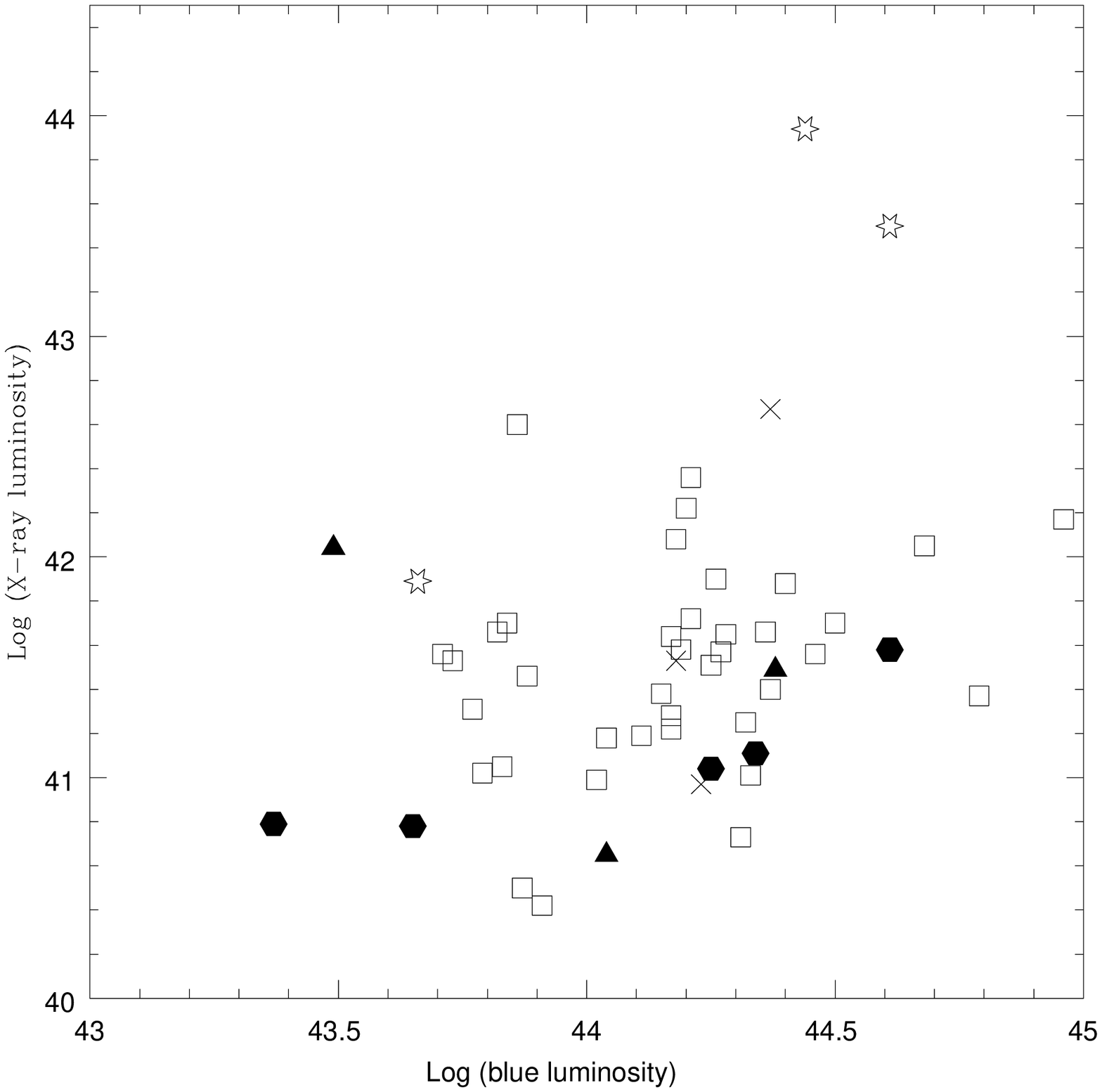}
\clearpage
\plotone{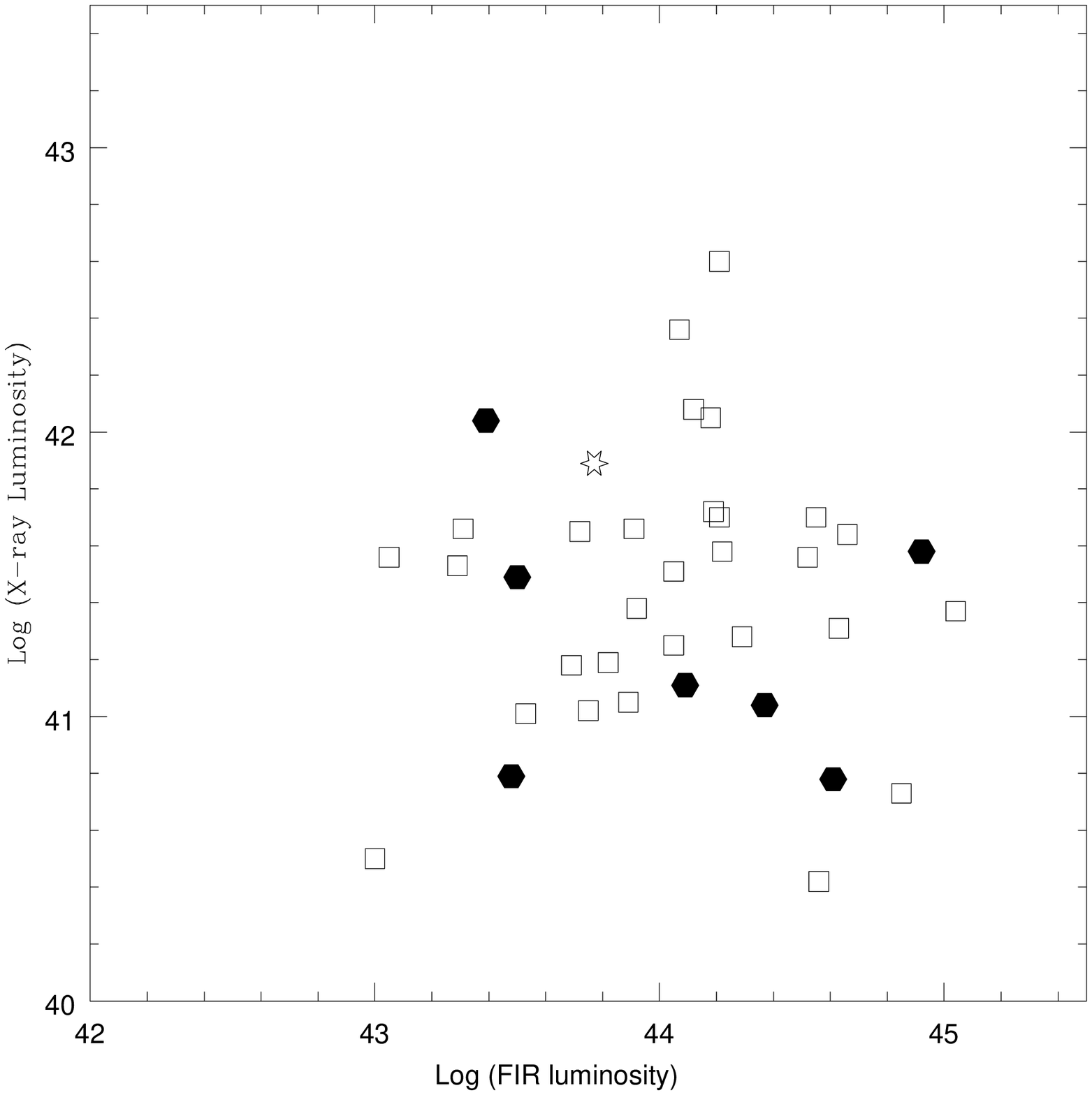}
\clearpage
\plotone{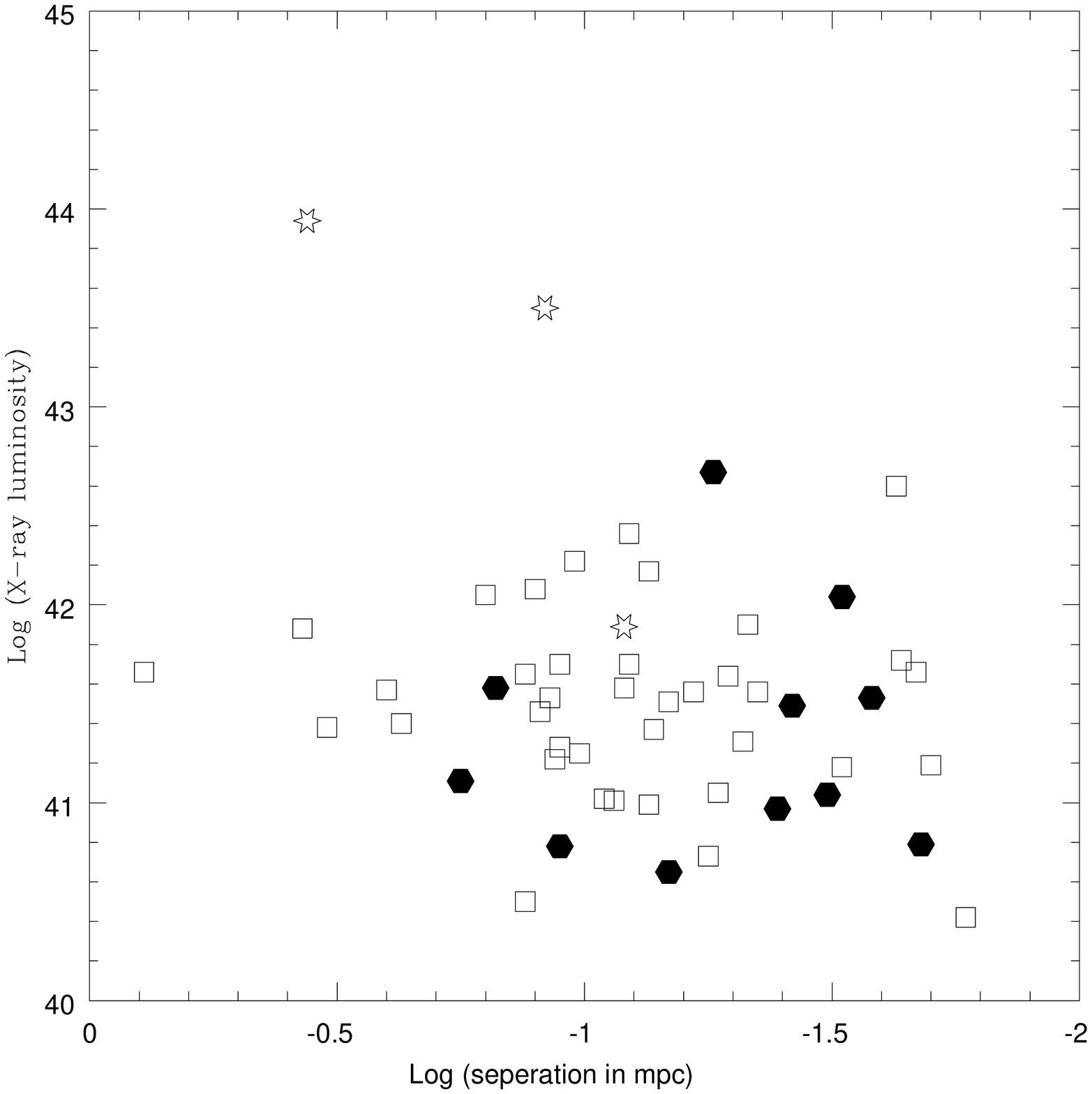}
\clearpage
\plotone{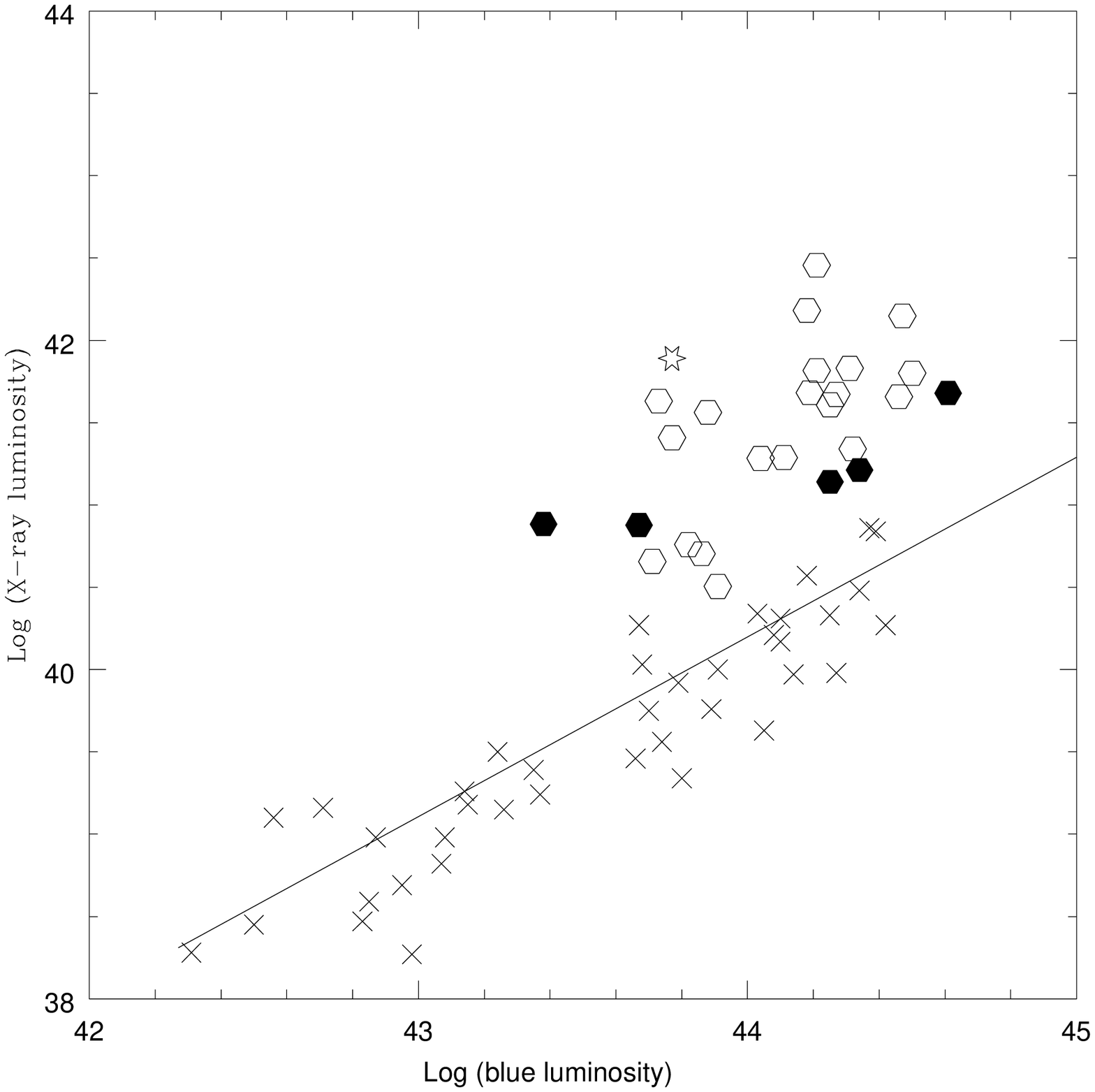}
\clearpage
\plotone{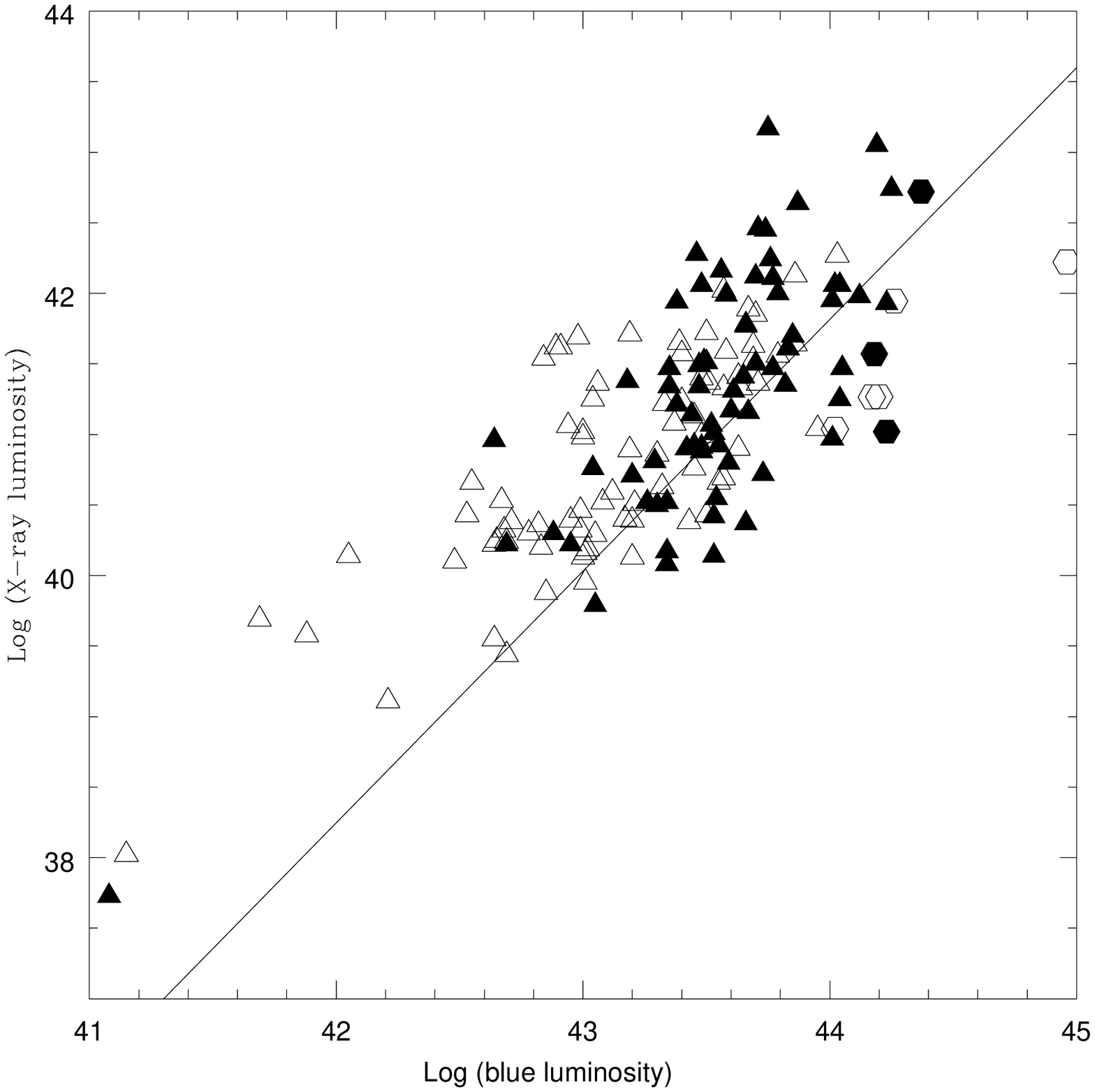}
\clearpage
\plotone{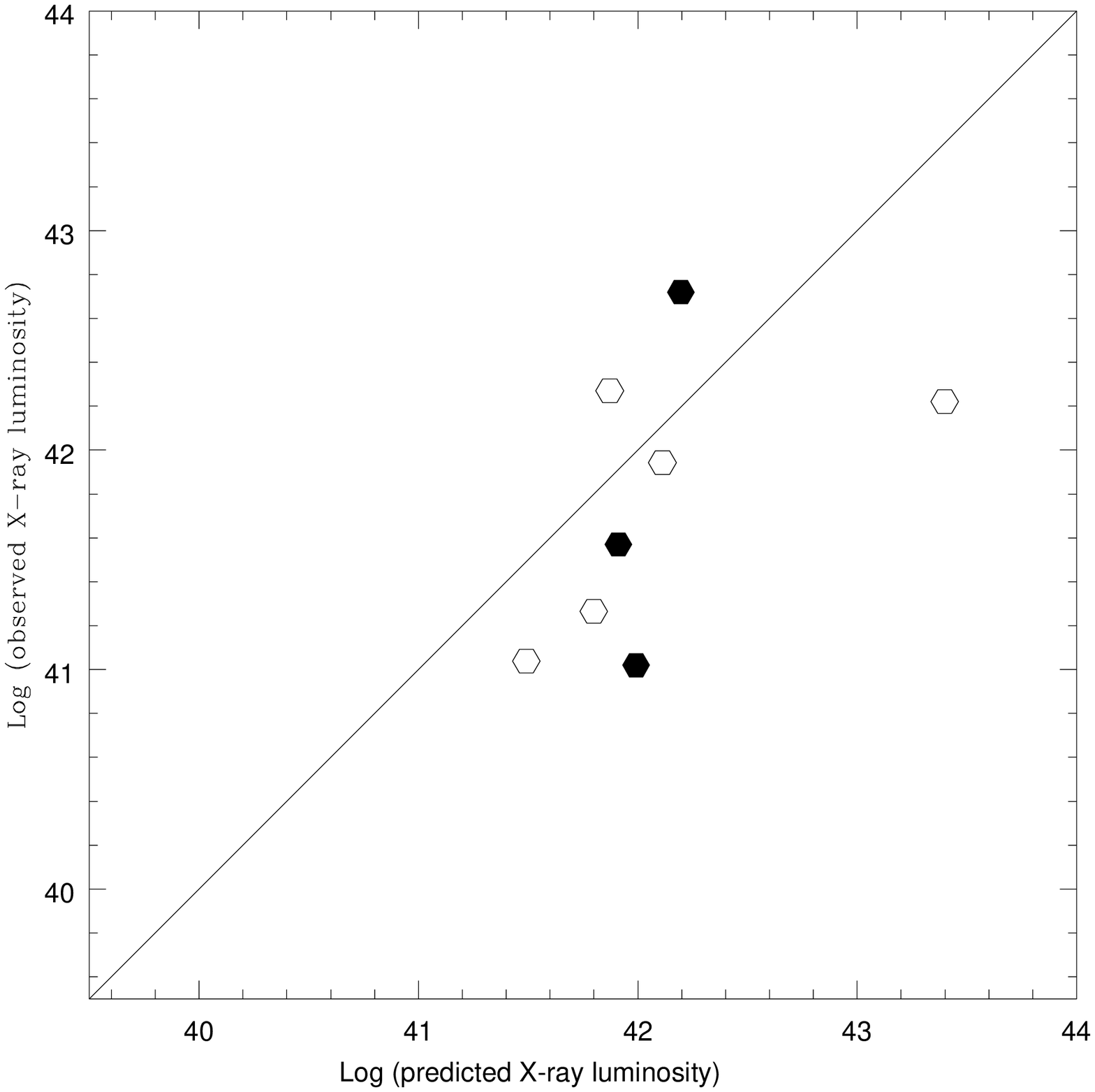}
\clearpage
\plotone{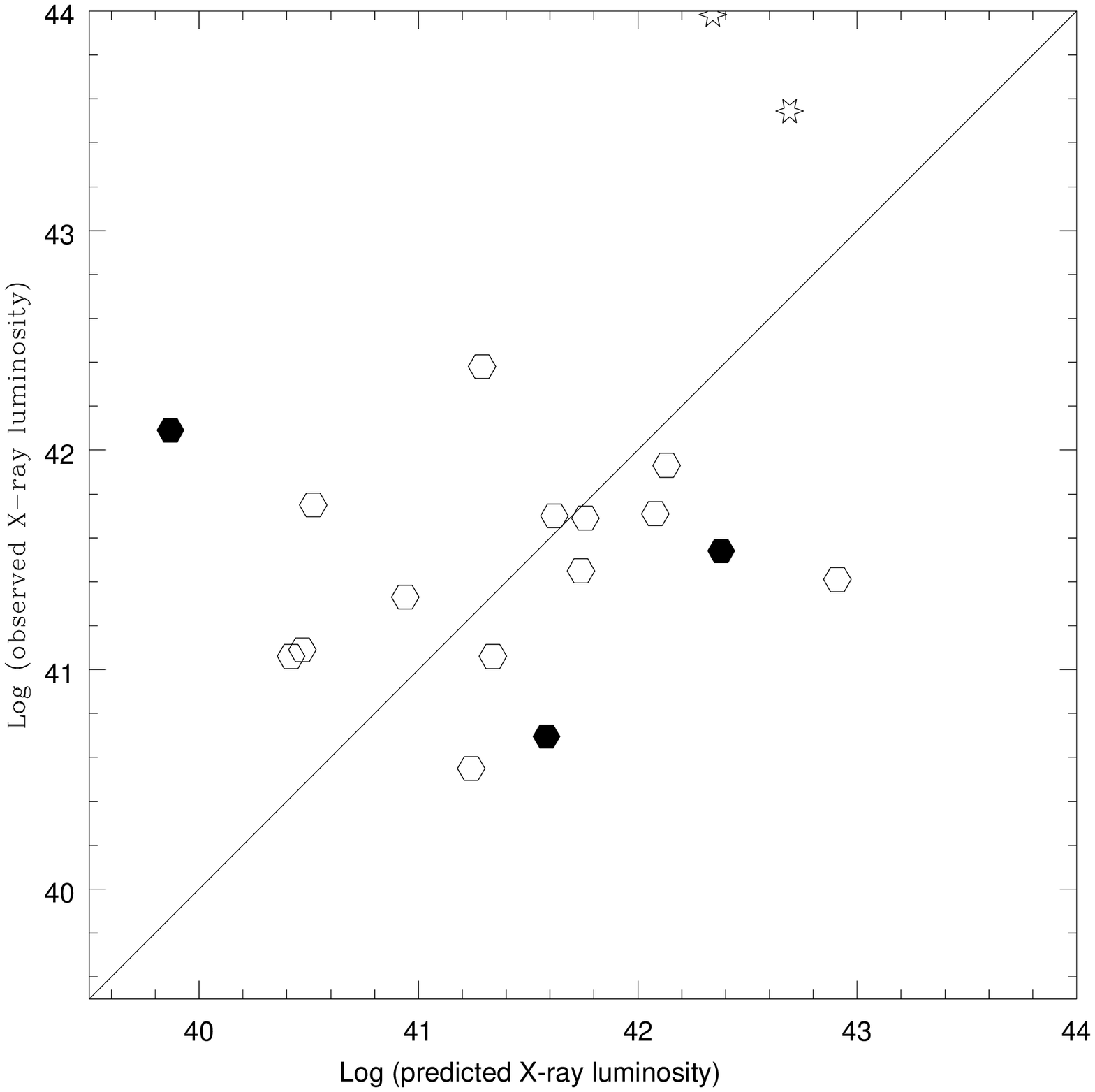}
\clearpage
\clearpage
\epsfxsize=2.0in \epsfbox{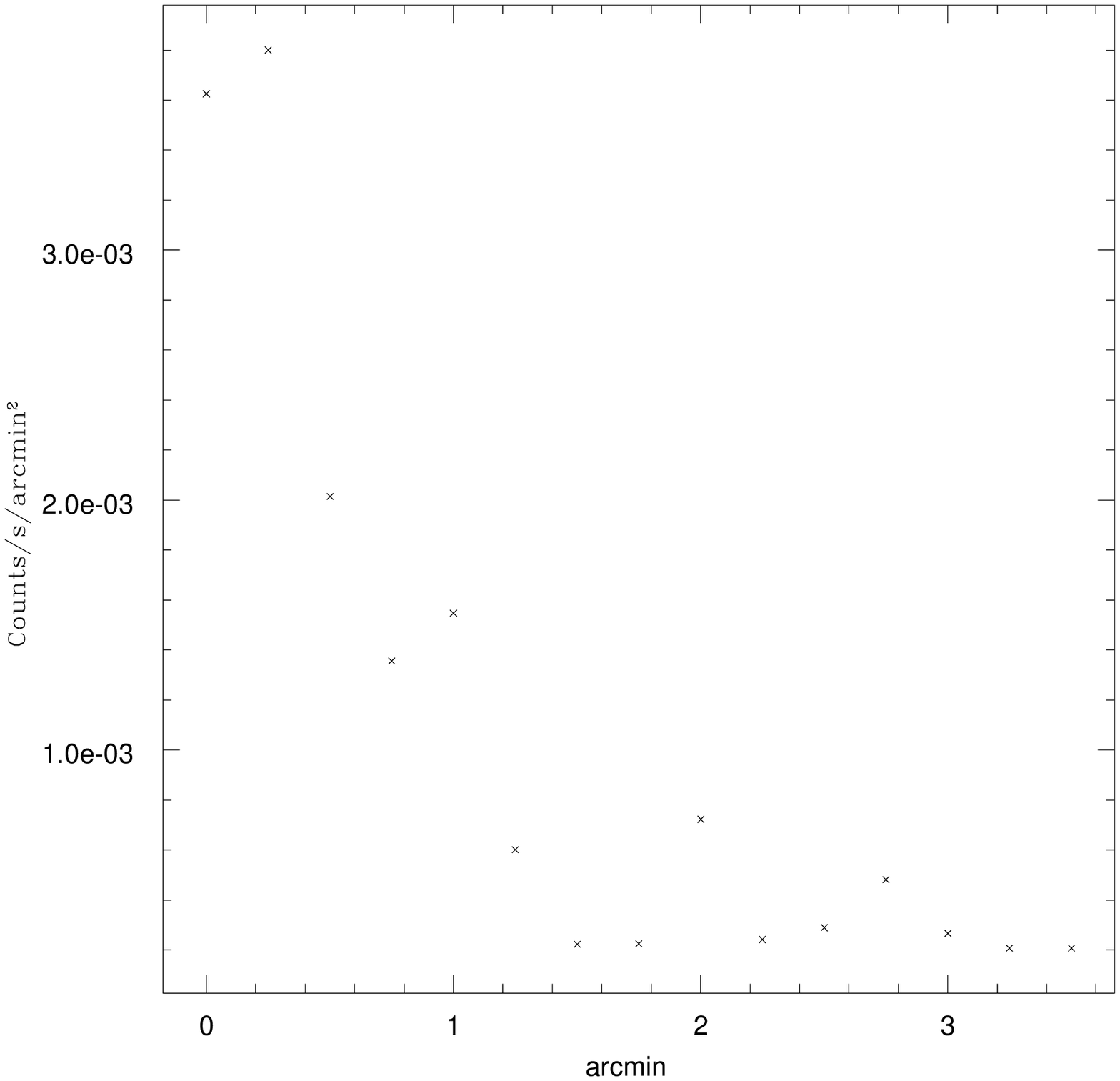}
\epsfxsize=2.0in \epsfbox{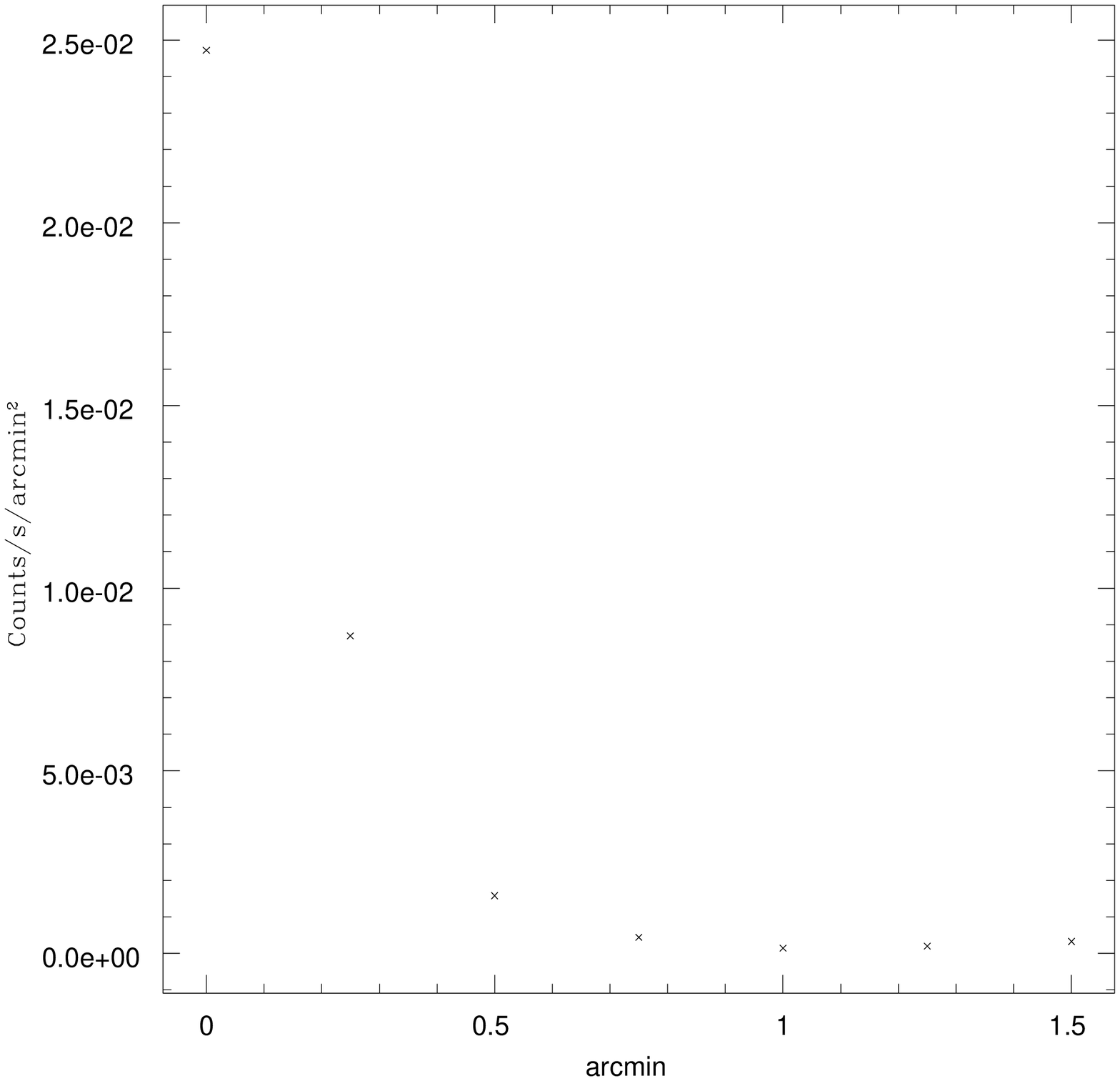}
\epsfxsize=2.0in \epsfbox{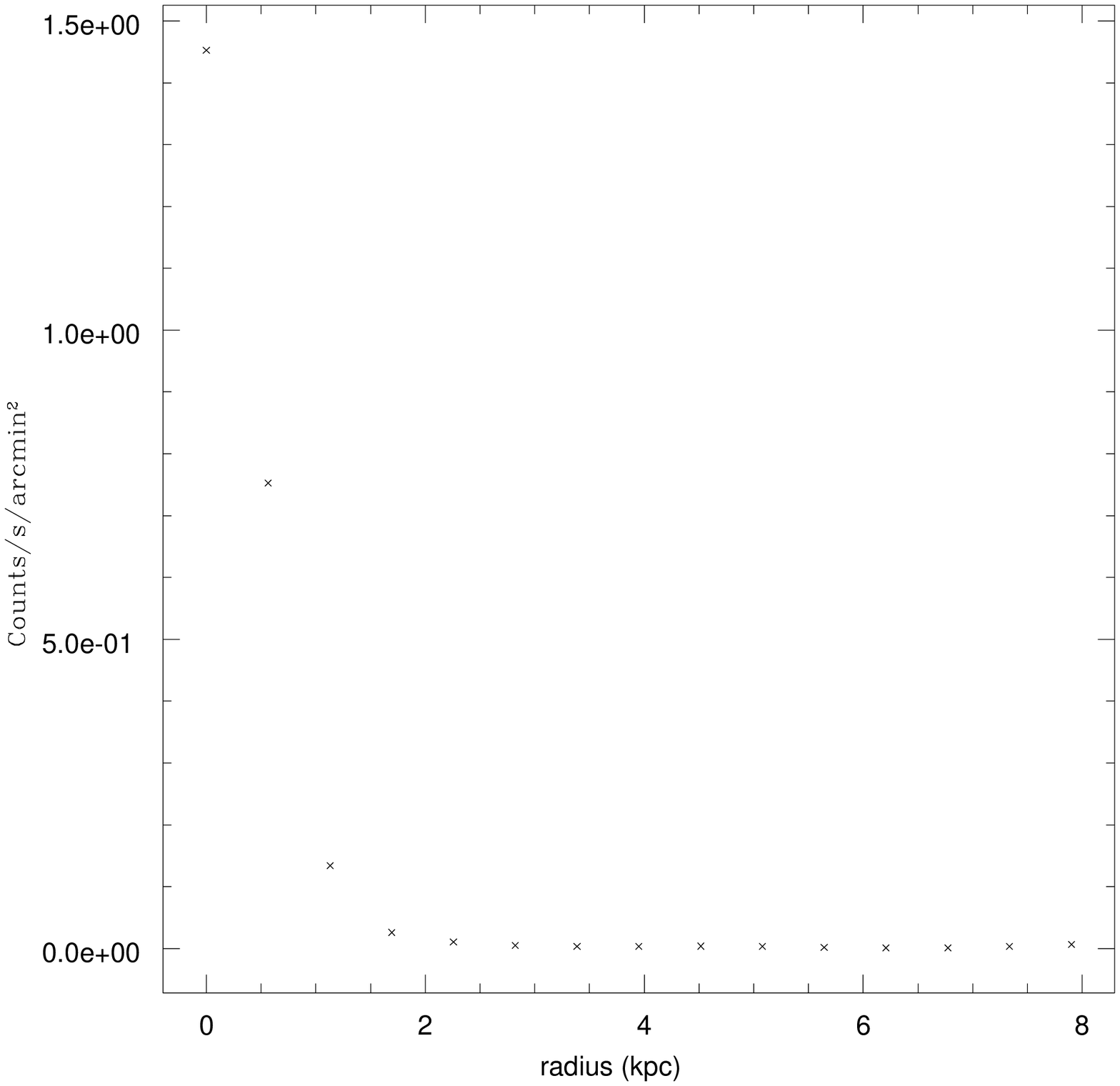}
\epsfxsize=2.0in \epsfbox{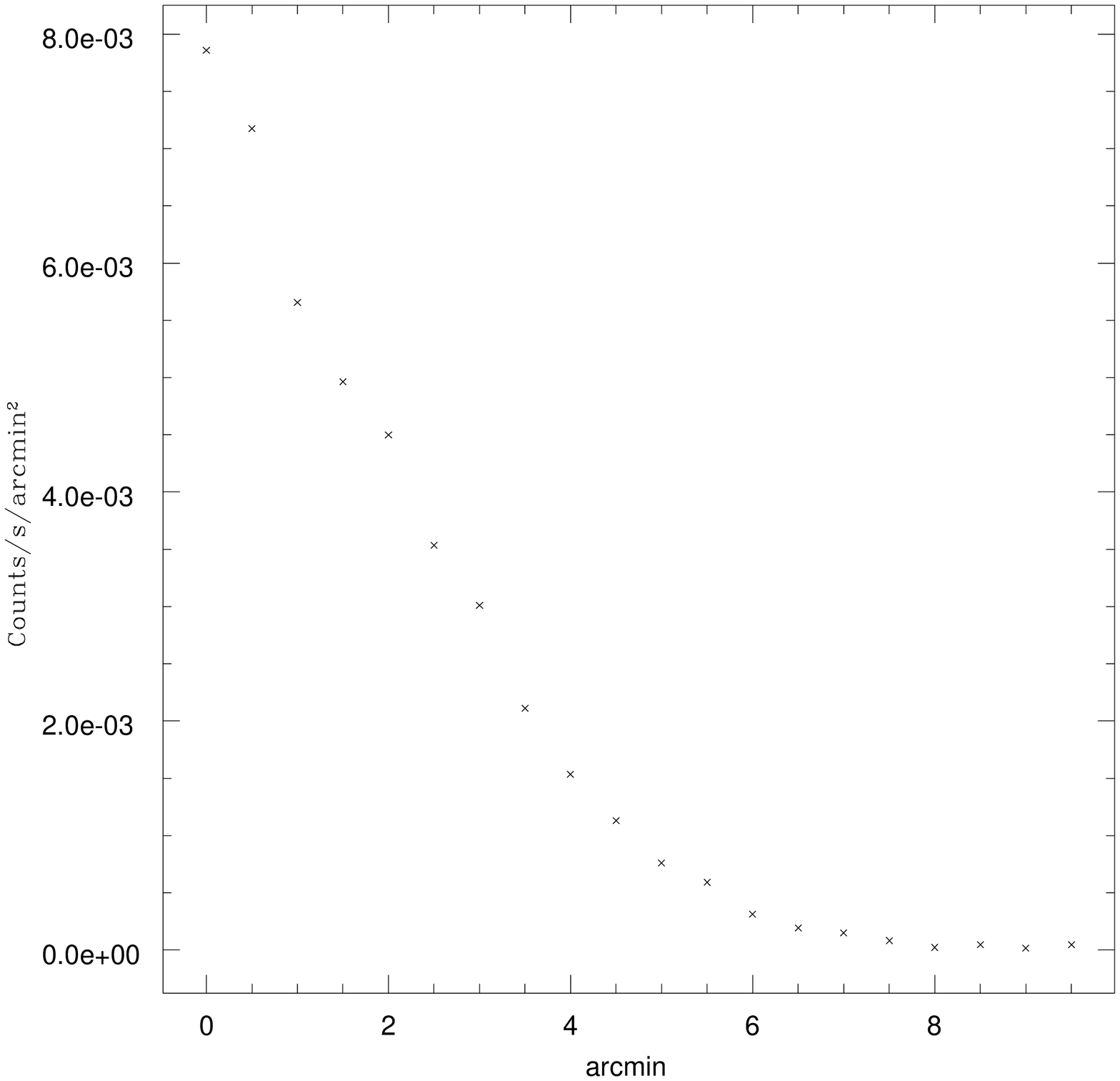}
\clearpage
\epsfxsize=2.0in \epsfbox{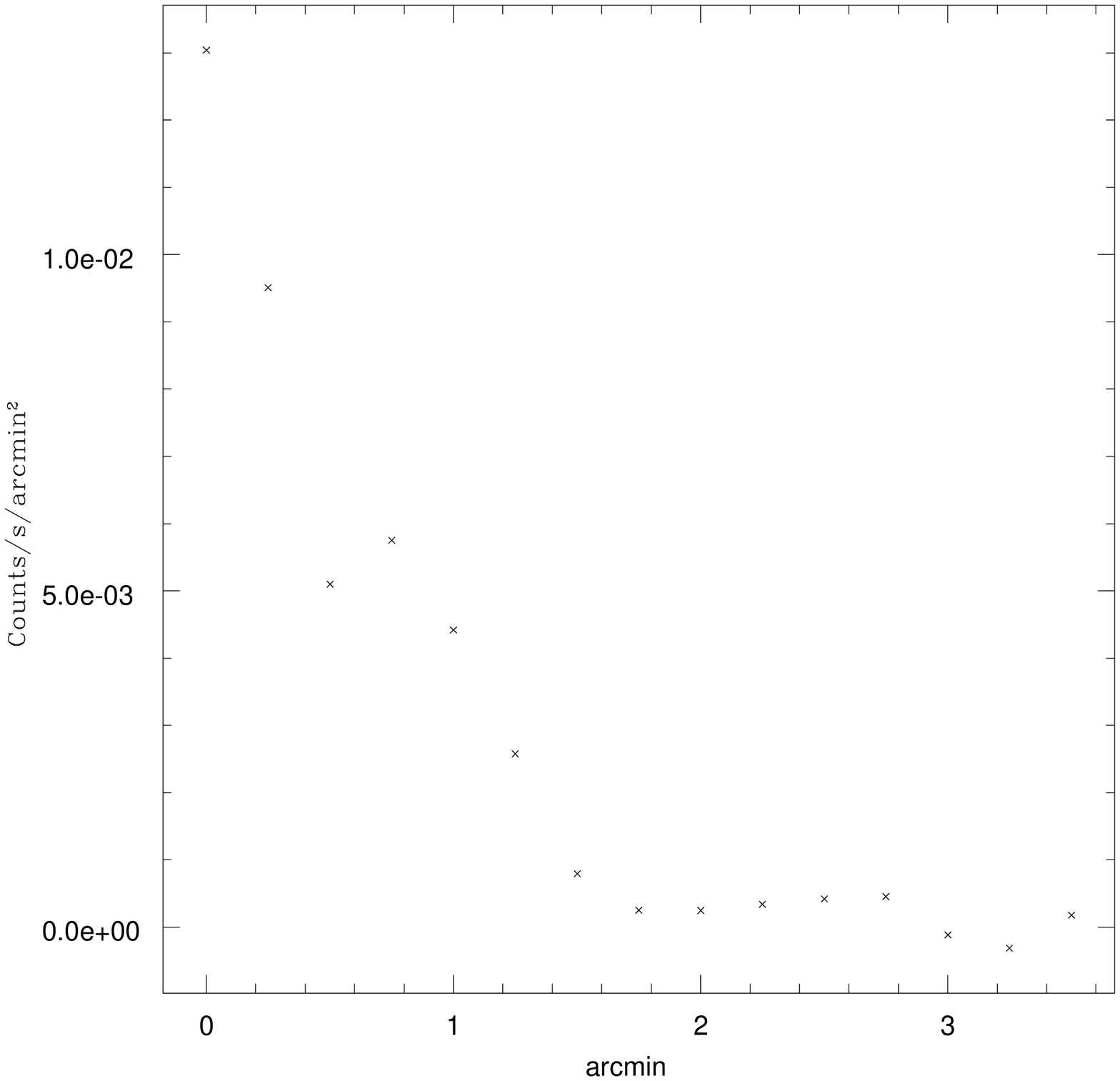}
\epsfxsize=2.0in \epsfbox{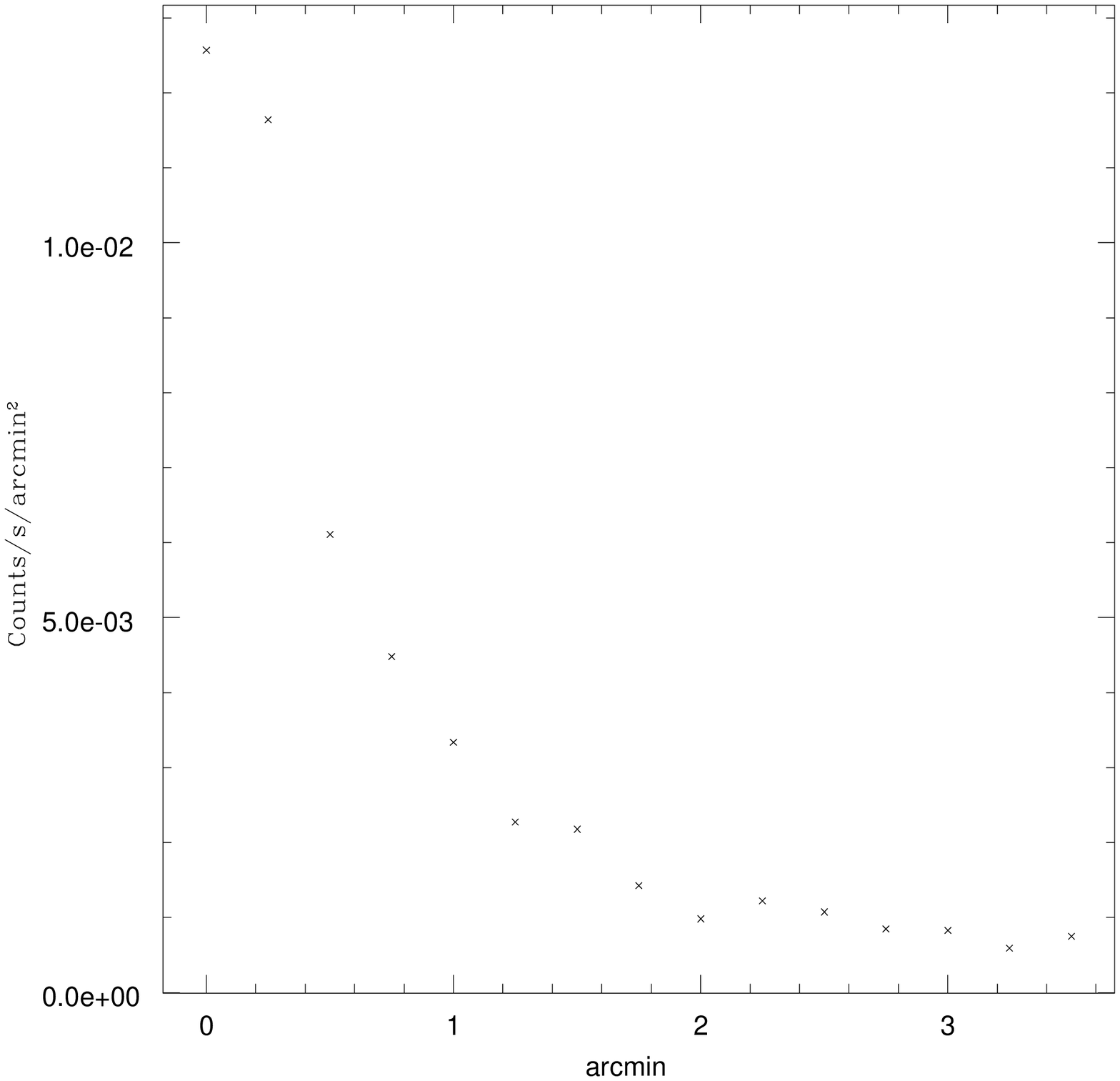}
\epsfxsize=2.0in \epsfbox{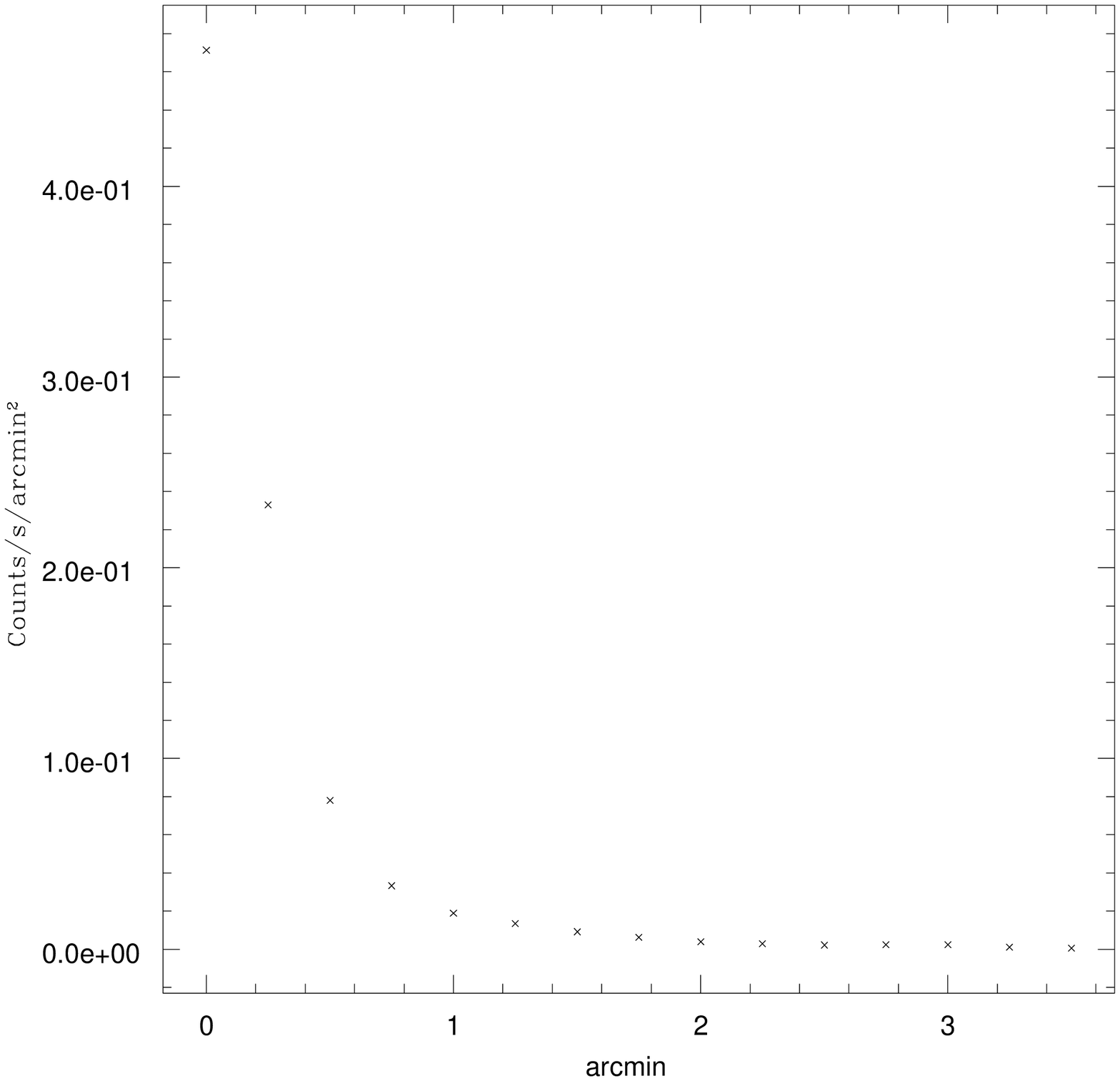}
\epsfxsize=2.0in \epsfbox{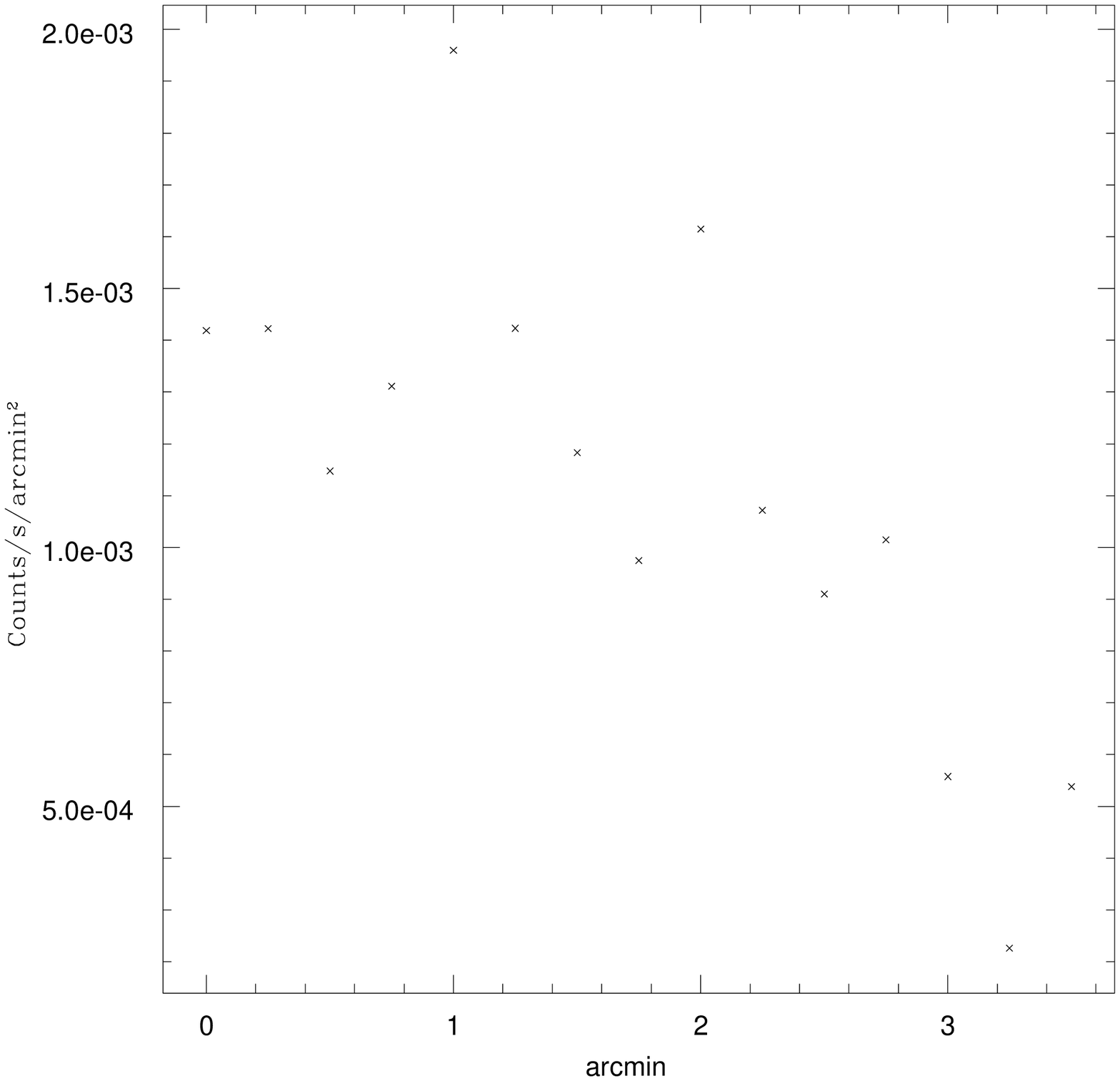}
\clearpage
\end{document}